\documentclass[prb,twocolumn,showpacs,preprintnumbers,amsmath,amssymb,superscriptaddress]{revtex4}
\usepackage{graphicx}
\usepackage{dcolumn}
\usepackage{bm}

\usepackage{epstopdf}

\usepackage{color}

\begin{document}

\title{Low temperature magnetic properties of  NpNi$_{5}$}

\author{A. Hen}\affiliation{European Commission, Joint Research Centre (JRC), Institute for Transuranium Elements (ITU), Postfach 2340, D-76125 Karlsruhe, Germany}\affiliation{Nuclear Engineering Department, Ben Gurion University, Beer-Sheva, Israel}
\author{E. Colineau}\affiliation{European Commission, Joint Research Centre (JRC), Institute for Transuranium Elements (ITU), Postfach 2340, D-76125 Karlsruhe, Germany}
\author{R. Eloirdi}\affiliation{European Commission, Joint Research Centre (JRC), Institute for Transuranium Elements (ITU), Postfach 2340, D-76125 Karlsruhe, Germany}
\author{J.-C. Griveau}\affiliation{European Commission, Joint Research Centre (JRC), Institute for Transuranium Elements (ITU), Postfach 2340, D-76125 Karlsruhe, Germany}
\author{N. Magnani}\affiliation{European Commission, Joint Research Centre (JRC), Institute for Transuranium Elements (ITU), Postfach 2340, D-76125 Karlsruhe, Germany}
\author{F. Wilhelm}
\affiliation{European Synchrotron Radiation Facility (ESRF), B.P.220, F-38043 Grenoble, France}
\author{A. Rogalev}
\affiliation{European Synchrotron Radiation Facility (ESRF), B.P.220, F-38043 Grenoble, France}
\author{J.-P. Sanchez}\affiliation{CEA, INAC-SPSMS, FR-38000 Grenoble, France}
\affiliation{Universit\'{e} de Grenoble Alpes, INAC-SPSMS, FR-38000 Grenoble, France}
\author{A. B. Shick}\affiliation{Institute of Physics, ASCR, Na Slovance 2, CZ-18221 Prague, Czech Republic}
\author{I. Halevy}\affiliation{Physics Department, Nuclear Research Center Negev, P.O. Box 9001, Beer-Sheva, Israel}\affiliation{Nuclear Engineering Department, Ben Gurion University, Beer-Sheva, Israel}
\author{I. Orion}\affiliation{Nuclear Engineering Department, Ben Gurion University, Beer-Sheva, Israel}
\author{R. Caciuffo}\affiliation{European Commission, Joint Research Centre (JRC), Institute for Transuranium Elements (ITU), Postfach 2340, D-76125 Karlsruhe, Germany}

\date{\today}

\begin{abstract}
We present the result of an extended experimental characterization of the hexagonal intermetallic Haucke compound NpNi$_{5}$. By combining macroscopic and shell-specific techniques, we determine the 5$f$-shell occupation number $n_f$ close to 4 for the Np ions, together with orbital and spin components of the ordered moment in the ferromagnetic phase below T$_C$ = 16 K ($\mu_{S}$ = -1.88~$\mu_{B}$ and $\mu_{L}$ = 3.91~$\mu_{B}$). The apparent coexistence of ordered and disordered phases observed in the M\"{o}ssbauer spectra is explained in terms of slow relaxation between the components of a quasi-triplet ground state. The ratio between the expectation value of the magnetic dipole operator and the spin magnetic moment ($3\langle T_{z}\rangle/ \langle S_{z}\rangle$ = +1.43) is positive and large, suggesting a localized character of the 5$f$ electrons. The angular part of the spin-orbit coupling ($\langle\vec{\ell}\cdot\vec{s}\rangle$ = -5.55) is close to the value of -6.25 calculated for trivalent Np ions in intermediate coupling approximation. The results are discussed against the prediction of first-principle electronic structure calculations based on the spin-polarized local spin density approximation plus Hubbard interaction, and of a mean field model taking into account crystal field and exchange interactions.
\end{abstract}


\maketitle

\section{Introduction}
The Haucke intermetallic compound NpNi$_{5}$ was discovered during systematic investigations of the neptunium-nickel phase diagram motivated by the search of optimal targets for minor actinides transmutation processes, and by concerns related to the safety of nuclear fuels interacting with cladding materials \cite{akabori97}. X-ray diffraction showed that NpNi$_{5}$ (contrary to the cubic UNi$_{5}$) has an hexagonal crystal symmetry and shares the CaCu$_{5}$ structure type with the Th and Pu analogs.  Based on the lattice parameters, the authors of Ref. [\onlinecite{akabori97}] derived a metallic radius of 0.1750 nm for the Np atoms in NpNi$_{5}$. According to Zachariasen \cite{zachariasen73}, such a value would correspond to a Np$^{4+}$ (5\textit{f}$^{3}$) oxidation state. Moreover, as the interatomic spacing of about 0.4 nm between Np atoms is substantially larger than the Hill limit (0.325 nm), the 5\textit{f} electrons should be localized \cite{hill71} and appreciable crystal field effects are expected.

More information on the physical properties of NpNi$_{5}$ is not available. On the other hand, the literature on the isostructural RNi$_{5}$ compounds (R = rare earth) is much wider. Several members of this system react reversibly with hydrogen and can absorb it in large quantities at ambient temperatures and pressures \cite{vanvucht70,kuijpers73,takeshita80,szajek03,yu08}. They have therefore been widely investigated for their potential as hydrogen and electrochemical storage material. Also the magnetic properties of RNi$_{5}$ compounds have attracted considerable attention, mainly because of their magnetocrystalline anisotropy, related to the interplay of crystal field (CF), spin-orbit, and exchange interactions acting on the R$^{3+}$ ions \cite{goremychkin85,barthem88,reiffers89,gubbens89,amato92,novak94,svoboda04}. In fact, as the 3\textit{d} band of Ni is almost filled by the rare earth valence electrons  \cite{buschow77},  the magnetic properties of RNi$_{5}$ are practically entirely due to the R ions. For non-magnetic rare earth (R = La, Ce, Lu, and Y), the compounds are enhanced Pauli paramagnets, otherwise they are ferromagnetic, with the highest T$_{C}$ = 32 K exhibited by GdNi$_{5}$ \cite{buschow77,gignoux76}. An exception is PrNi$_{5}$, which is characterized by a well-isolated singlet CF ground state, and is therefore a Van Vleck paramagnet \cite{folle81,andres77,kuchin04,kuchin06}.

Here, we present the results obtained by Superconducting Quantum Interference Device (SQUID) magnetometry, specific heat, M\"{o}ssbauer spectroscopy, and x-ray magnetic circular dichroism (XMCD) measurements carried out on NpNi$_{5}$ polycrystalline samples. We show that the compound orders ferromagnetically below T$_{C}$ = 16 K, with an ordered moment $\mu_{0}$ $\sim$ 2 $\mu_{B}$ carried out by Np$^{3+}$ ions and a very small contribution ($\sim$ 0.08 $\mu_{B}$ per Ni atom) from the Ni sublattice. Information on the low-energy part of the electronic spectrum is obtained from the analysis of the specific heat curve. A combined analysis of the different experimental data sets, allowed us to determine key parameters associated with the electronic structure of the system. These results are used to benchmark the predictions of first principle electronic structure calculation based on the spin-polarized local spin density approximation plus Hubbard interaction.

The remainder of this paper is structured as follows. In
Sec.\ II, we provide details on synthesis and experimental procedures;  in Sec.\ III we present the experimental results and compare them with the outcome of mean-field calculations. The results of first principle calculations are presented in Sec. \ IV. Discussion and conclusions are found in Sec.\ V. and \ VI.

\section{Experimental details}
Polycrystalline samples of NpNi$_{5}$ were prepared by arc melting stoichiometric amounts of high-purity elemental constituents (99.9 \% Np, 99.999 \% Ni) on a water-cooled copper hearth, under Ar (6N) atmosphere. A Zr alloy was used as an oxygen getter. The weight losses were examined after arc melting and resulted to be less than 0.5 $\%$. The sample was melted 5 times and crushed before the last melt, to ensure complete homogeneity of the alloy button.

The same procedure has been followed to prepare a sample of the non-magnetic iso-structural compound ThNi$_{5}$, which has been used to determine the vibrational contribution to the specific heat of NpNi$_{5}$.

Crystallographic phase analyses were performed at room temperature by x-ray diffraction on samples with a mass of about 25 mg, finely ground and dispersed on a Si wafer. Data were collected in reflection mode with a Bruker D8 diffractometer installed inside a $\alpha-\gamma$ glove box, using Cu-K$\alpha_{1}$ radiation ($\lambda$ = 0.15406 nm) selected by a Ge (111) monochromator. A 1-D position sensitive detector was used to cover the angular range from 15 to 120 degrees, with incremental steps of 0.0085 degrees.

dc-magnetization and magnetic susceptibility measurements were carried out in the temperature range 2$-$300 K, with an external magnetic field up to 7 T on a 196 mg sample using the MPMS-7 SQUID from Quantum Design (QD). The specific heat experiments were performed using the relaxation method on a 10.9 mg polycrystalline sample using the QD PPMS-14 platform in the temperature range 2.2$-$300 K and in a magnetic field up to 14 T. For these measurements, the sample was enrobed in Stycast$\circledR$ 1266 epoxy encapsulant, to prevent external radioactive contamination. The resin contribution to the measured heat capacity was subtracted according to a standard procedure.  The $^{237}$Np M\"{o}ssbauer measurements were performed in transmission geometry on a powder absorber with a thickness of 140 mg cm$^{-2}$ of Np. The M\"{o}ssbauer source ($\sim$108 mCi of $^{241}$Am metal) was kept at 4.2 K, while the temperature of the absorber was varied from 4.2 to 140 K in discreet steps. The spectra were recorded with a sinusoidal drive system using conventional methods. The velocity scale was calibrated with reference to a NpAl$_{2}$ standard (B$_{hf}$ = 330 T at 4.2 K). No contributions associated with the small amount of NpO$_{2}$ impurity phase have been observed in the measured quantities. In particular no anomaly was visible around 25 K, the temperature below which long-range order of magnetic triakontadipoles is stabilized in NpO$_{2}$.\cite{santini06,magnani08}

The x-ray-absorption-spectroscopy (XAS) and x-ray magnetic circular dichroism (XMCD) experiments were carried out at the European Synchrotron Radiation Facility (ESRF) using the ID12 beamline, which is dedicated to polarization-dependent spectroscopy in the photon-energy range from 2 to 15 keV. Data were collected at the M$_{4,5}$ edges of Np (3.6-3.9 keV). After monochromatization with a double-crystal-Si(111), the rate of circular polarization of the photons provided by the Helios-II helical undulator (in excess of 0.95) was reduced to about 0.42 at the M$_{5}$ and 0.50 at the M$_{4}$ absorption edge.  The x-ray-absorption spectra were recorded using the total-fluorescence-yield detection mode in backscattering geometry for parallel $\mu^{+}$(E) and antiparallel $\mu^{-}$(E) alignments of the photon helicity with respect to a 7 T external magnetic field applied along the beam direction. In order to get XMCD spectra free of experimental artifacts, measurements were also performed for the opposite direction of the applied magnetic field. For these measurements, we used a 3.5 mg sample glued with a conductive resin (Dupont 4929) inside an Al capsule, with a 100-$\mu$m-thick Be window and a 13-$\mu$m-thick kapton foil for protection.

\section{Experimental results}
\subsection{Crystal structure}
\begin{figure}
\includegraphics[width=8.0cm]{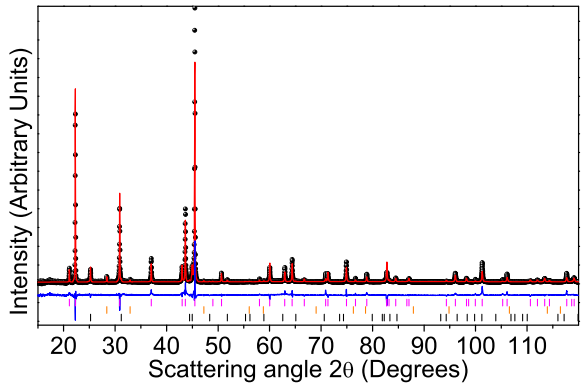}
\caption{Color online. Observed (dots) and calculated (red line) x-ray diffraction pattern recorded at room temperature for NpNi$_{5}$. The lower trace (blue line) is the difference profile. The intensity distribution is plotted as a function of the full diffraction angle 2$\theta$ (Cu K$\alpha_{1}$ radiation). Vertical ticks indicate calculated angular positions of the Bragg peaks for (upper row, magenta) the NpNi$_{5}$ phase,  (middle row, orange) for an impurity phase (NpO$_{2}$, $\sim$ 3 $\%$ in weight), and (lower row, black) a SiO$_{2}$ contamination from the quartz mortar.
\label{diffpattNpNi5}}
\end{figure}

The room temperature powder diffraction pattern of NpNi$_{5}$ is shown in Fig.~\ref{diffpattNpNi5}. The Rietveld analysis was performed with the X'Pert HighScore Plus software package of PANalytical. The background was fitted with a 6$^{th}$-order polynomial, and the shape of the Bragg peaks was described by a pseudo-Voigt function. The observed Bragg peaks can be indexed in the hexagonal $P6/mmm$ space group, with lattice parameters a = 0.48501(5) nm and c = 0.39841(2) nm, in very good agreement with previous estimates \cite{akabori97}. Additional observed Bragg peaks reveal the presence of NpO$_{2}$ as an impurity phase ($\sim$ 3\% in weight), and some SiO$_{2}$ from the agate mortar used to grind the sample. No other impurities were observed.

\begin{figure}
\includegraphics[width=8.0cm]{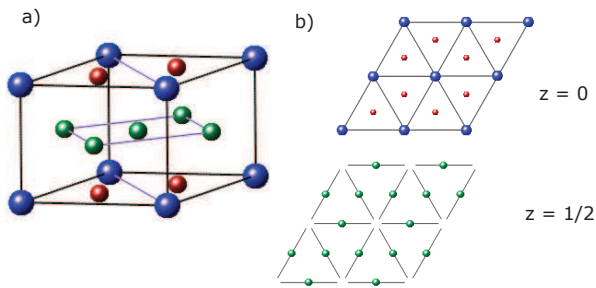}
\caption{Color online; (a) crystallographic unit cell of the Haucke hexagonal intermetallic compound NpNi$_{5}$ (Space Group $P6/mmm$). Np atoms (blue, large spheres) occupy the $1a$ Wyckoff position at $0~0~0$ with point symmetry $6/mmm$, whereas Ni atoms are located at the $2c$ ($1/3~2/3~0$, $\bar{6}m2$, red spheres) and $3g$ ($1/2~0~1/2$, $mmm$, green spheres) special positions. (b) Atom arrangements on the $z = 0$ and $z = 1/2$ planes (2$\times$2 unit cells). \label{crysstrucNpNi5}}
\end{figure}

The best fit was obtained assuming the CaCu$_{5}$-type structure typical of Haucke compounds, with full occupation of $1a$, $2c$, and $3g$ sites and fixed  isotropic temperature factors B (B$_{Np}$ = 0.2 {\AA}$^{2}$; B$_{Ni}$ = 0.4 {\AA}$^{2}$). Any attempt to vary the occupancy of additional sites resulted in an increase of the weighted-pattern index (R$_{wp}$ = 0.13). The crystal structure is shown in Fig.~\ref{crysstrucNpNi5}. The z = 0 plane contains Np and Ni atoms located, respectively, at the $1a$ and $2c$ special positions, whilst the plane at z = 1/2 contains Ni atoms centered at the $3g$ Wyckoff sites. The first, second, and third neighbors of Np are 6 Ni ($2c$), 12 Ni ($3g$), and 2 Np, respectively. The smallest nearest Np-Np distance is of 0.39841(5) nm, well above the so-called Hill limit of $\sim$0.32 nm that separate Np compounds with localized $5f$ states from itinerant systems. Appreciable crystal field effects are therefore expected.

\subsection{Magnetic measurements}
\begin{figure}
\includegraphics[width=8.0cm]{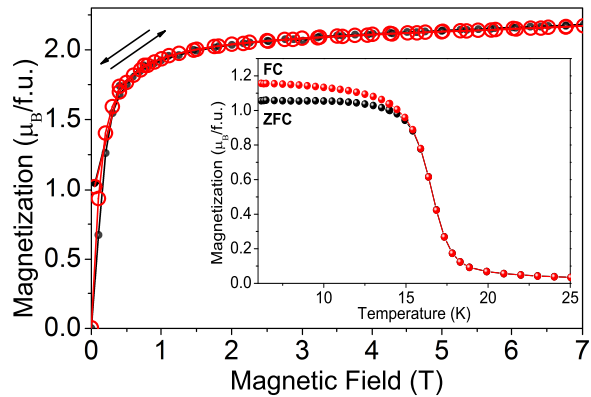}
\caption{Color online. Magnetic field dependence of the magnetization measured on a polycrystalline sample of NpNi$_{5}$ at 2 K (black dots) and 5 K (red open circles). The inset shows the temperature dependence of the magnetization in a field $\mu_{0}$H = 0.1 T under field cooled (FC) and zero-field cooled (ZFC) conditions. \label{magnetizationNpNi5}}
\end{figure}

\begin{figure}
\includegraphics[width=8.0cm]{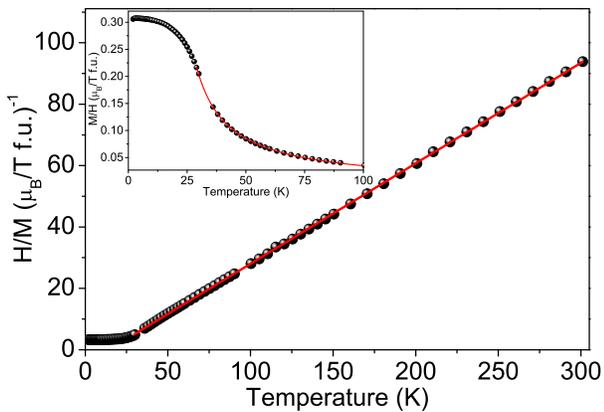}
\caption{Temperature dependence of the NpNi$_{5}$ inverse magnetic susceptibility $H/M$, measured in a field $\mu_{0}$H = 7 T. The straight line is a fit to the Curie-Weiss law, giving an effective paramagnetic moment $\mu_{eff}$ = 3.69 $\mu_{B}$ and a Curie-Weiss temperature $\theta$ = 14.7 K.  The temperature dependence of the susceptibility is shown in the inset. \label{invsusceNpNi5}}
\end{figure}

The magnetic field ($\mu_{0}H$) dependence of the magnetization $M(T, \mu_{0}H)$ is shown in Fig.~\ref{magnetizationNpNi5} for temperatures $T$ equal to 2 and 5 K. The $T$ dependence of  $M(T, \mu_{0}H)$, measured after field-cooling (FC) and zero-field-cooling (ZFC) conditions in a field $\mu_{0}H$ = 0.1 T is shown in the inset. The observed magnetization curve is typical for a soft ferromagnet with vanishing coercive field and a Curie temperature T$_{C} \simeq$ 16 K, below which irreversibility effects associated to domain walls motions are observed. Below T$_{C}$, the magnetization saturates around $\mu_{0}H$ = 2 T, reaching a value of $\sim$2.2  $\mu_{B}/f.u.$ at $\mu_{0}H$ = 7 T. The temperature dependence of the inverse magnetic susceptibility $H/M$, measured up to 300 K in a field of 7 T, is shown in Fig.~\ref{invsusceNpNi5}. Above $\sim$~40 K, it can be fitted by the Curie-Weiss law with an effective paramagnetic moment  $\mu_{eff}$ = 3.69 $\mu_{B}$ and a Curie-Weiss temperature $\theta_{CW}$ = 14.7 K.
The effective moment is very close to that of a Np$^{4+}$ free-ion in intermediate coupling (3.8 $\mu_B$), whereas it is much larger than for a trivalent ion (2.88 $\mu_B$). However, the latter case
cannot be excluded since the fact that the  \emph{ordered} moment on the Ni sites is very small (as determined in the next subsection) does not imply a negligible contribution to the \emph{effective} moment. This was clearly shown in the case of NdNi$_{5}$ where the much larger than free ion value can be ascribed to the Ni contribution \cite{barthem89}.

\subsection{Specific heat measurements}
\begin{figure}
\includegraphics[width=8.0cm]{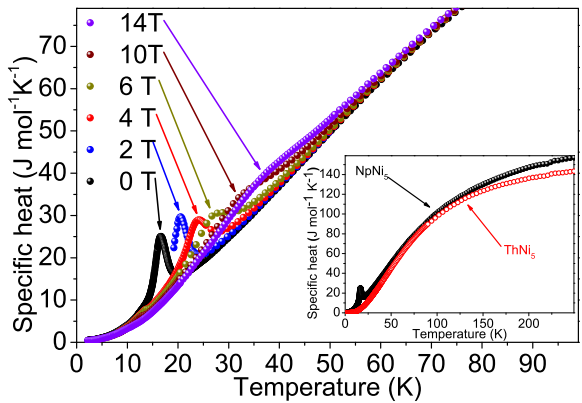}
\caption{Temperature dependence of the specific heat measured for NpNi$_{5}$ in an external magnetic field with amplitude between 0 and 14 T. The specific heat curves measured from 2 to 300 K for (black circle) NpNi$_{5}$ and (red open circles) ThNi$_{5}$ are shown in the inset. \label{CpNpNi5}}
\end{figure}

Figure \ref{CpNpNi5} shows the experimental data for both NpNi$_5$, and ThNi$_5$. The anomaly observed at T$_{C}$ for the former broadens and moves towards higher temperatures upon application of an external magnetic field, as expected for an anisotropic polycrystalline ferromagnetic sample.

\begin{figure}
\includegraphics[width=8.0cm]{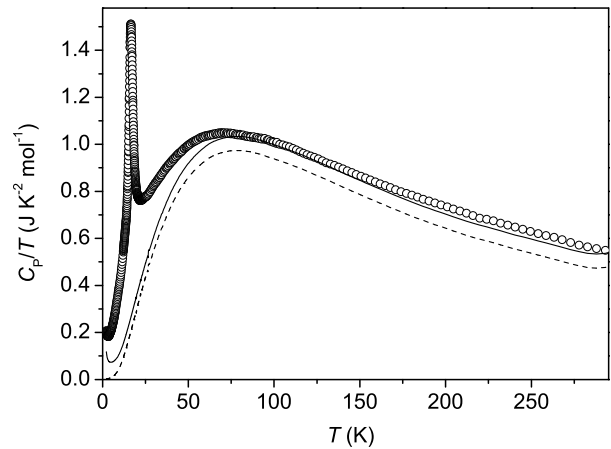}
\caption{Temperature dependence of the $C_{p}/T$ ratio plotted as a function of T.
The dashed line is the vibrational contribution estimated from the measurements performed on the isostructural compound ThNi$_5$.
The full line is the sum of the vibrational contribution, a constant electronic term $\gamma = 0.06$ J K$^{-2}$ mol$^{-1}$, and a
nuclear Schottky contribution expressed as $C_N/T^3$ with $C_N=0.625$ J K/mol.
\label{CpT}}
\end{figure}

\begin{figure}
\includegraphics[width=8.0cm]{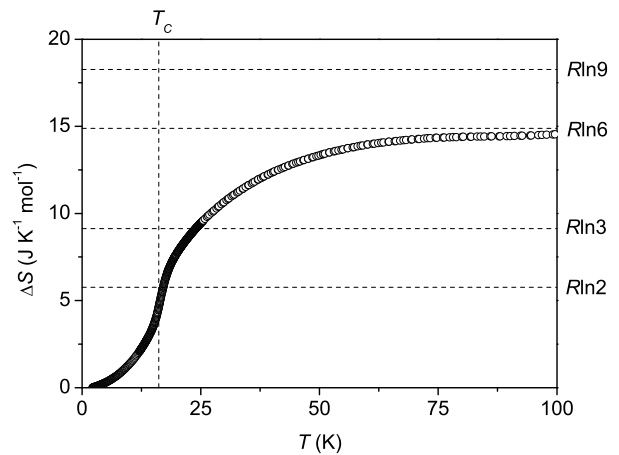}
\caption{Temperature dependence of the magnetic entropy S extracted by numerically integrating the $C_{p}/T$ data with the background subtracted.
\label{S(T)}}
\end{figure}

The vibrational contribution has been estimated by subtracting a constant electronic term $\gamma = 0.037$ J K$^{-2}$ mol$^{-1}$ from the $C_p/T$ measurements performed on ThNi$_5$, and
is compared to the NpNi$_5$ data in Fig. \ref{CpT}.
The full background used in order to extract the magnetic contribution to the specific heat of NpNi$_5$
is the sum of the vibrational contribution, a constant electronic term $\gamma = 0.06$ J K$^{-2}$ mol$^{-1}$, and a
nuclear Schottky contribution expressed as $C_N/T^3$, with $C_N=0.625$ J K/mol calculated from the M\"{o}ssbauer splitting at 4 K. This has been subtracted to the $C_p/T$ data of NpNi$_5$ before integrating the result
in order to obtain the temperature dependence of the magnetic entropy change (Fig. \ref{S(T)}). Its behavior is qualitatively in line with what expected for
Np$^{3+}$: at the jump around 16 K the entropy change is roughly equal to $R$ln2, indicating a doublet ground state.
The full magnetic entropy for the lowest $J$ manifold is recovered around room temperature.

\subsection{M\"{o}ssbauer spectroscopy}
\begin{figure}
\includegraphics[width=8.0cm]{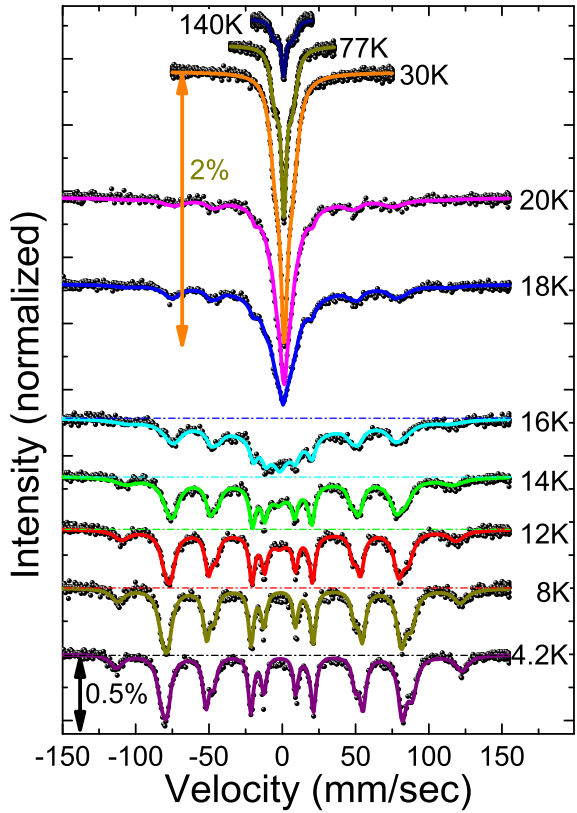}
\caption{Temperature evolution of the M\"{o}ssbauer spectra measured for NpNi$_{5}$. Dots are measured data, solid lines are the results of the fit to the model described in the text. Horizontal dashed lines indicate the baseline for each spectrum. \label{MossNpNi5}}
\end{figure}

$^{237}$Np M\"{o}ssbauer spectra recorded for NpNi$_{5}$ at different temperatures, from 4.2 to 140 K, are shown in Fig.~\ref{MossNpNi5}. A broadening of the central absorption line below 25 K seems to signal the occurrence of critical fluctuations, whereas a full magnetic splitting is observed below 14 K. Spectra calculated by solving the complete Hamiltonian for the hyperfine interactions, with a Lorentzian shape for the absorption lines, are shown by solid lines. The slight asymmetry of the magnetic split spectra is typical of systems dominated by magnetic interactions in presence of weaker quadrupolar ones. A good fit is obtained by considering a single set of hyperfine parameters, which is consistent with the crystal structure of the compound. The main component of the electric field gradient ($eq \equiv V_{zz}$) is taken collinear with the magnetization, and the asymmetry parameter ($\eta$) is set to zero, because of the axial symmetry of the Np site.

The observed isomer shift is $\delta_{IS}$ = -12.2 mm/s with reference to NpAl$_{2}$, compatible with a 5$f^{4}$ trivalent state of Np in a metallic environment \cite{halevy12}. At 4.2 K the hyperfine field is $B_{hf}$ = 437 T, corresponding to an ordered magnetic moment at the Np site  $\mu_{Np}$ = 2.03(5) $\mu_{B}$ ($B_{hf}$/$\mu_{Np}$ =215 $\pm$ 5 T/$\mu_{B}$), \cite{dunlap85} and the quadrupole coupling constant is $e^{2}qQ \simeq$ -10.5 mm/s. From the magnetization saturation value of 2.2 $\mu_{B}$/f.u., assuming as a crude approximation that the contribution of the conduction band is -10$\%$ of the total magnetization, the average magnetic moment per Ni atom in NpNi$_{5}$ is roughly estimated as $\sim$ 0.08 $\mu_{B}$.

\begin{figure}
\includegraphics[width=8.0cm]{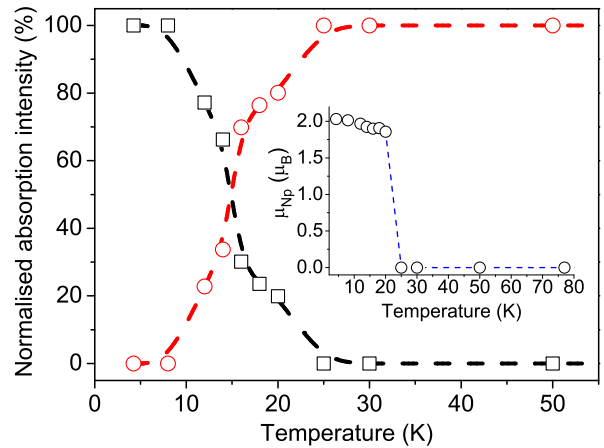}
\caption{Color online. Temperature dependence of the relative integrated  M\"{o}ssbauer absorption intensities associated to the sextet (open black squares) and to the central absorption line (open red circles). The inset shows the temperature variation of the Np ordered moment, as deduced from the hyperfine field. Dashed lines are a guide to the eye. \label{phaseratio}}
\end{figure}

 As shown in Fig.~\ref{phaseratio}, the temperature dependence of the order parameter shows an abrupt step at $T_C$ rather than a continuous decrease. At the same time, the M\"{o}ssbauer absorption spectra show the persistence of both a broad central absorption peak below $T_C$ and of a magnetic splitting above. These observations could indicate a first order nature of the phase transition. However, the absence of hysteresis in the magnetic and specific heat measurements, as well as their shape, do not support this conclusion. A different explanation involves crystal field effects as considered in the following sections.

\subsection{Crystal-field model}
Given that the nature of 5$f$ electrons in NpNi$_5$ is expected to be essentially localized, we have attempted to model its properties by simple crystal-field (CF) calculations.
The CF Hamiltonian appropriate for the $D_{6h}$ symmetry of the Np sites is
\begin{equation}
H_{CF} = B_{2}^{0} O_{2}^{0} + B_{4}^{0} O_{4}^{0} + B_{6}^{0} O_{6}^{0} + B_{6}^{6} O_{6}^{6}
\label{Hcf}
\end{equation}
and reasonable values of the $B_k^q$ parameters can be obtained by rescaling those determined for the isostructural PrNi$_5$ compound \cite{andres79}.
Table \ref{tab:Np3vsNp4CF} gives the crystal-field splittings and the wavefunctions calculated for the lowest $J$ multiplets of the
$5f^4$ (Np$^{3+}$, $J=4$) configuration in the paramagnetic phase.

\begin{table}
\caption{\label{tab:Np3vsNp4CF} Crystal-field energy spectra in the paramagnetic phase calculated for Np$^{3+}$ in NpNi$_5$. In the ordered phase the ground state becomes a singlet with composition $ 0.927 \left| + 4 \right\rangle -0.375 \left| - 2 \right\rangle $.}
\begin{ruledtabular}
\begin{tabular}{rcc}
\multicolumn{1}{c}{$E$ (K)} & \multicolumn{1}{c}{symmetry} & \multicolumn{1}{c}{wavefunction} \\
\hline
0 & $\Gamma_5^{(1)}$ & $ 0.917 \left| \pm 4 \right\rangle -0.400 \left| \mp 2 \right\rangle $ \\
21 & $\Gamma_4$ & $ 1/\sqrt{2} \left| 3 \right\rangle - 1/\sqrt{2} \left| -3 \right\rangle $ \\
193 & $\Gamma_1$ & $ \left| 0 \right\rangle $ \\
290 & $\Gamma_6$ & $ \left| \pm 1 \right\rangle $  \\
487 & $\Gamma_5^{(2)}$ & $ 0.917 \left| \pm 2 \right\rangle -0.400 \left| \mp 4 \right\rangle $  \\
492 & $\Gamma_3$ & $ 1/\sqrt{2} \left| 3 \right\rangle + 1/\sqrt{2} \left| -3 \right\rangle $ \\
\end{tabular}
\end{ruledtabular}
\end{table}

Adding the mean-field term
\begin{equation}
H_{MF} = \lambda g^{2} \mu_{B}^{2} \left[ -J_{z}\langle J_{z} \rangle +
\langle J_{z} \rangle ^2 /2 \right]
\label{Hmf}
\end{equation}
with $\lambda = 6.8 T/\mu_{B}$ triggers a second-order transition to a ferromagnetic state with the easy magnetization direction along the $c$ axis,
and reproduces exactly both the experimentally determined Curie temperature (T$_C$ = 16 K) and the ordered moment (2.03 $\mu_B$).

\subsection{XMCD spectra}
XMCD is associated with time-reversal symmetry breaking by a magnetic field and involves electric dipole or electric quadrupole transitions promoting an electron in a spin-orbit split core state to an empty valence state of the absorbing atom. The technique provides an element- and shell-specific probe for studying the electronic structure of a wide range of materials \cite{vanderlaan13}.  In this experiment, XMCD spectra have been measured at
the $M_{4,5}$ ($3d_{3/2,5/2} \rightarrow 5f$) Np absorption edges.

\begin{figure}
\includegraphics[width=8.0cm]{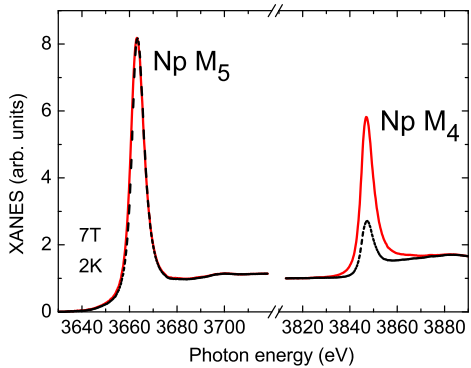}
\caption{Color online. NpNi$_{5}$ x-ray absorption spectra recorded
at the M$_{5}$ and M$_{4}$ Np edges for parallel ($\mu^{+}$(E), dashed black line) and antiparallel ($\mu^{-}$(E), solid red line) alignments of the photon helicity with respect to a 7 T external magnetic field applied along the beam direction (sample temperature 2 K). The spectra have
been corrected for self-absorption effects and incomplete circular polarization of the incident beam. \label{XANESNpNi5}}
\end{figure}

The XAS spectra, measured with the sample kept at 2 K in a magnetic field of 7 T, are shown in Fig. \ref{XANESNpNi5} for both for both right- and left- circularly polarized x-rays. The spectra have been corrected for self-absorption effects and for incomplete circular polarization rates of incident x-ray photons. The self-absorption corrections were done in the semi-infinite sample approximation, taking into account the chemical composition, the density, the background contributions (fluorescence of subshells and matrix, coherent and incoherent scattering), and the geometrical factors \cite{goulon82,troger92,pfalzer99}.
The Np edge-jump intensity ratio M$_{5}$/M$_{4}$ has been normalized to 1.57, the value reported in the XCOM tables by Berger $\textit{et al.}$ in Ref. \onlinecite{berger},
which is close to the $1:2/3$ statistical edge-jump ratio (defined as the ratio between the occupation numbers for the two spin-orbit-split core levels j = 3/2 and 5/2).

\begin{figure}
\includegraphics[width=8.0cm]{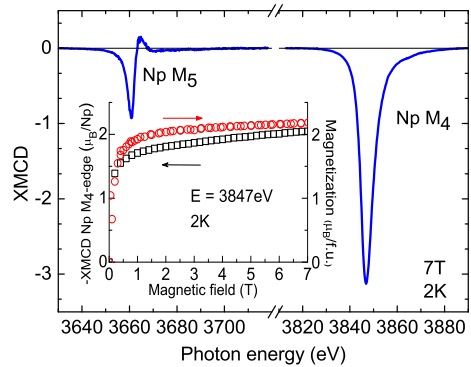}
\caption{NpNi$_{5}$ XMCD spectra measured at the M$_{5}$ and M$_{4}$ Np edges in an applied field
of 7 T at 2 K. The spectra have been corrected for self-absorption
effects and for the incomplete circular polarization rate. Inset: (black squares) element-specific XMCD magnetization curve recorded at the maximum XMCD signal at the Np M$_{4}$ absorption
edge (E = 3847 eV) with the sample kept at 2 K, normalized to the ordered magnetic moment obtained by M\"{o}ssbauer spectroscopy; (red circles) SQUID magnetization curve measured at 2 K. \label{XMCDNpNi5}}
\end{figure}

The XMCD signal has been obtained as the difference
of the corrected x-ray-absorption spectra, $\Delta I_{M_{4,5}} =
\mu^+(E) - \mu^-(E)$, whereas the white line intensities $I_{M_{4,5}} $ were obtained after a subtraction of a step function convoluted
by a Voigt line shape, in order to remove contributions
due to transitions into the continuum.
 The results are shown in Fig.~\ref{XMCDNpNi5}, together with the field dependence of the dichroic signal at the $M_{4}$ edge, normalized at 7 T to  the value of the Np ordered magnetic moment deduced from M\"{o}ssbauer spectroscopy. This provides a Np-specific magnetization curve, which is compared with the magnetization measured by SQUID at 2 K. The difference between the two data sets provides the conduction electrons and Ni sublattice contributions to the total magnetization.

The orbital and spin contributions to the total magnetic moment carried by the Np atoms can be separately determined by appropriate sum rules \cite{vanderlaan96,vanderlaan04}. In particular,  the ground state expectation value of the orbital moment is obtained from the  total dichroic signal, $\Delta I_{M_{5}}+\Delta I_{M_{4}}$, as \cite{thole92}:

\begin{equation}\label{om}
\langle L_{z}\rangle = \frac{n_{h}}{I_{M_{5}}+I_{M_{4}}} (\Delta I_{M_{5}}+\Delta I_{M_{4}})
\end{equation}

where $I_{M_{4,5}}$ is the integrated intensity of the isotropic x-ray absorption spectrum at the Np $M_{4,5}$ edges, and $n_{h}$ is the number of holes in the $5f$ shell. The spin polarization $\langle S_{z}\rangle$ is instead obtained from a second sum rule, stating that \cite{carra93}:

\begin{equation}\label{sm}
\langle S_{z}\rangle + 3\langle T_{z}\rangle = \frac{n_{h}}{2(I_{M_{5}}+I_{M_{4}})} (\Delta I_{M_{5}}-\frac{3}{2}\Delta I_{M_{4}})
\end{equation}

where the so-called magnetic dipole term $T_{z}$ is the component along the quantization axis of an operator associated to charge and magnetic anisotropy  \cite{collins95} and correlating spin $\vec{s}_{i}$ and position $\vec{r}_{i}$ of individual electrons, $\vec{T} = \sum_{i} [\vec{s}_{i} -3(\vec{r}_{i} \cdot \vec{s}_{i})/r_{i}^{2}]$.
Furthermore, the expectation value of the angular part of the valence states spin-orbit operator, $\langle\psi|\vec{\ell}\cdot\vec{s}|\psi\rangle$, can be obtained from the XAS branching ratio, $B = I_{M_{5}}/(I_{M_{5}}+I_{M_{4}})$, as \cite{thole88}

\begin{equation}\label{so}
\frac{2\langle\vec{\ell}\cdot\vec{s}\rangle}{3n_{h}}-\Delta = -\frac{5}{2}(B-\frac{3}{5}),
\end{equation}

where $\Delta$ is a quantity dependent from the electronic configuration\cite{vanderlaan04}
 (for Np$^{3+}$, $\Delta$ = -0.005)
and the experiment provides an isotropic branching ratio $B$ = 0.746(5).
Assuming  $n_{h}$ = 10, from Eq.~\ref{so} one obtains $\langle\vec{\ell}\cdot\vec{s}\rangle$ = -5.55, a value close to the one (-6.25) calculated in the intermediate coupling approximation for Np$^{3+}$ free ions \cite{vanderlaan96}. Morevover, as $\langle \psi|\vec{\ell}\cdot\vec{s}|\psi\rangle$ = $3n^{5f}_{7/2}/2$ - $2n^{5f}_{5/2}$, one obtains for the occupation numbers of the $j$ = 5/2 and $j$ = 7/2 5$f$ sub-shells $n^{5f}_{5/2}$ = 3.3 and $n^{5f}_{7/2}$ = 0.7 ($n^{5f}_{5/2}$+$n^{5f}_{7/2}$ = 4).

Inserting in Eq.~\ref{om} the experimental values $\Delta I_{M_{5}}$ = -4.3037, $\Delta I_{M_{4}}$ = -27.1271, and $I_{M_{5}}+I_{M_{4}}$ = 80.4663, one obtains $\langle L_{z}\rangle$ = -3.91, and therefore $\mu_{L}$ = 3.91 $\mu_{B}$. The total Np moment, as given by the M\"{o}ssbauer hyperfine field, is $\mu_{Np}$ = $\mu_{L}$+$\mu_{S}$ = 2.03 $\mu_{B}$, and one obtains $\mu_{S}$ = -1.88 $\mu_{B}$ and $\langle S_{z}\rangle$ = 0.94.

Table \ref{tab:Np3vsNp4XMCD} lists some parameters and observables obtained from XMCD data assuming a trivalent oxidation state for neptunium, compared
with the same quantities calculated for the ground state in the ordered phase by the mean-field model outlined in the previous subsection.

From Eqs.~\ref{om} and~\ref{sm} one obtains $(\langle S_{z}\rangle + 3\langle T_{z}\rangle)$ = 2.26. This corresponds to $\langle T_{z}\rangle$ = 0.44 and to a ratio $3\langle T_{z}\rangle/ \langle S_{z}\rangle$ = +1.43, which differ by about 5$\%$ from the value of +1.36 obtained by similar experiments in Np$_{2}$Co$_{17}$ \cite{halevy12}, which is also a localized system, but is about 70$\%$ larger than the value (0.83) found for the itinerant ferromagnet NpOs$_{2}$ \cite{wilhelm13}.

\begin{table}
\caption{\label{tab:Np3vsNp4XMCD} Comparison between moments deduced from combining XMCD and M\"{o}ssbauer results and corresponding values calculated with the mean-field model for Np$^{3+}$ ions, as described in the text.}
\begin{ruledtabular}
\begin{tabular}{ccc}
 & obs. & calc. \\
\hline
$\mu_{\rm ord}$ ($\mu_B$) & 2.03 & 2.03  \\
$\langle L_\zeta \rangle$ & -3.91 & -4.27  \\
$\langle S_\zeta \rangle + 3 \langle T_\zeta \rangle$ & 2.26 & 1.75 \\
$\langle \vec{\ell} \cdot \vec{s} \rangle$ & -5.55 & -6.26 \\
\end{tabular}
\end{ruledtabular}
\end{table}

\section{First-principle calculations}
To examine theoretically the electronic structure of NpNi$_{5}$
and to make a comparison with experimental data, we
performed  spin-polarized local spin density approximation (LSDA) as well as
LSDA plus Hubbard U (LSDA+U) calculations, using the experimental lattice parameters listed above.
All calculations are performed making use of the in-house
implementation~\cite{shick97,shick01} of the full-potential
linearized augmented plane wave (FP-LAPW) method. This FP-LAPW
version includes all relativistic effects: scalar-relativistic and
spin-orbit coupling (SOC). The radii of the atomic muffin-tin (MT)
spheres are set to 2.8~a.u.~(Np) and 2.1~a.u.~(Ni). The basis set
size is characterized by the parameter
$R_{Np} \times K_{\rm max}=8.4$ and the
Brillouin zone is sampled with  765 $k$~points.

First, we apply the conventional spin-polarized LSDA, assuming
anti-parallel coupling between Ni-3$d$ and Np-5$f$ spin moments
\cite{brooks89}. Table ~\ref{tab:lsda} reports the calculated spin
($\mu_S$), orbital ($\mu_L$), and total ($\mu = \mu_S +\mu_L$)
magnetic moments in the "muffin-tin" spheres around Np and Ni atoms
(in $\mu_B$ units), together with the total values (including
contributions from the interstitial region) per formula unit (f.u.)
of $\mu_S$, $\mu_L$ and $\mu$.  It is seen that LSDA yields
substantially smaller value of the magnetization/f.u. (0.76
$\mu_B$), than experimentally observed (2.2 $\mu_B$).

Next, we apply LSDA+U calculations, making use of  relativistic
(including SOC) "around-mean-field" (AMF)-LSDA+U,
Ref.~[\onlinecite{shick97}].  The Coulomb interaction in the Np
5$f$~shell is parameterized by Slater integrals $F_0= 3.00$~eV,
$F_2=7.43$~eV, $F_4=4.83$~eV and $F_6=3.53$~eV, as given in Ref.
~[\onlinecite{moore09}]. They correspond to commonly accepted values
$U = 3$~eV and $J = 0.6$~eV for the Coulomb and exchange
interactions parameters, respectively.

\begin{table}[htbp]
\caption{\label{tab:lsda}  Spin ($\mu_S$), orbital ($\mu_L$),
and total ($\mu = \mu_S +\mu_L$) magnetic moments in the ``muffin-tin'' spheres around Np and Ni atoms
(in $\mu_B$ units), together with total values (including interstitial contributions) per formula unit (f.u.) of $\mu_S$, $\mu_L$ and $\mu$.}

\begin{tabular}{ccccccccccc}
  \hline
&&\multicolumn{3}{c}{LSDA(+SOC)}& &\multicolumn{3}{c}{AMF-LSDA+U} \\
  \hline
Atom& Site& $\mu_S$& $\mu_L$& $\mu$& &$\mu_S$  & $\mu_L$ & $\mu$& $\mu_{exp}$ \\
\hline
  Np             & $1a$& -2.89 & 2.92 &-0.03 && -1.63 & 3.93& 2.30& 2.03\\
  Ni$_{1}$ & $2c$ &  0.20 &-0.04 &0.20  && 0.11 & -0.04&0.07& 0.08 \footnote{Average obtained by dividing the Ni sublattice moment by the number of sites.}\\
  Ni$_{2}$ & $3g$&   0.23& 0.00 &0.23 &&  0.19 & 0.00& 0.19& 0.08 $^{a}$\\
  \hline
  \multicolumn{2}{c}{Total} & -2.08& 2.84 & 0.76 && -0.96 &3.81&2.85& 2.2 \footnote{From magnetization data.}\\
\hline
\end{tabular}
\end{table}

\begin{figure}
\includegraphics[width=7cm]{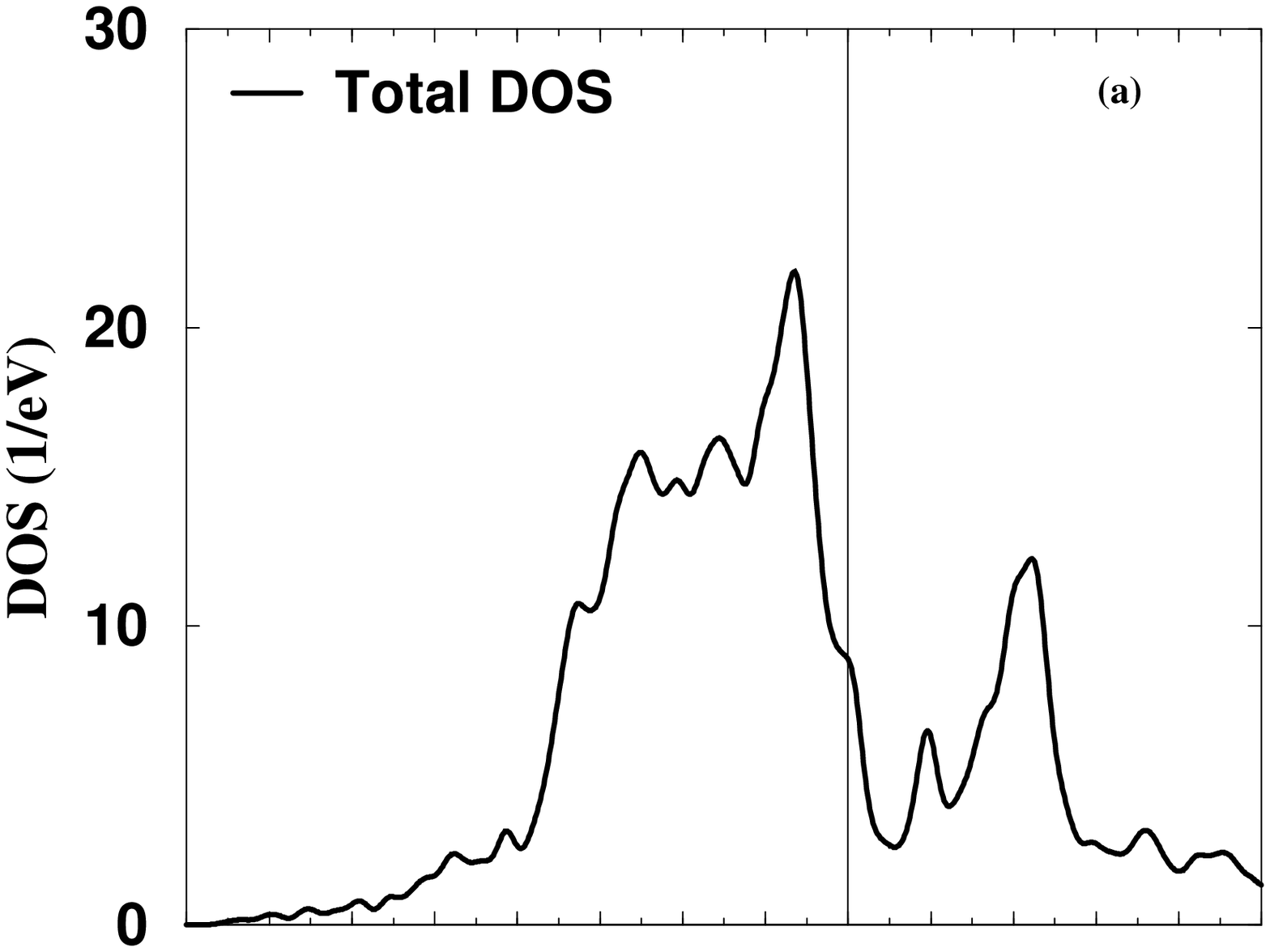}
\includegraphics[width=7cm]{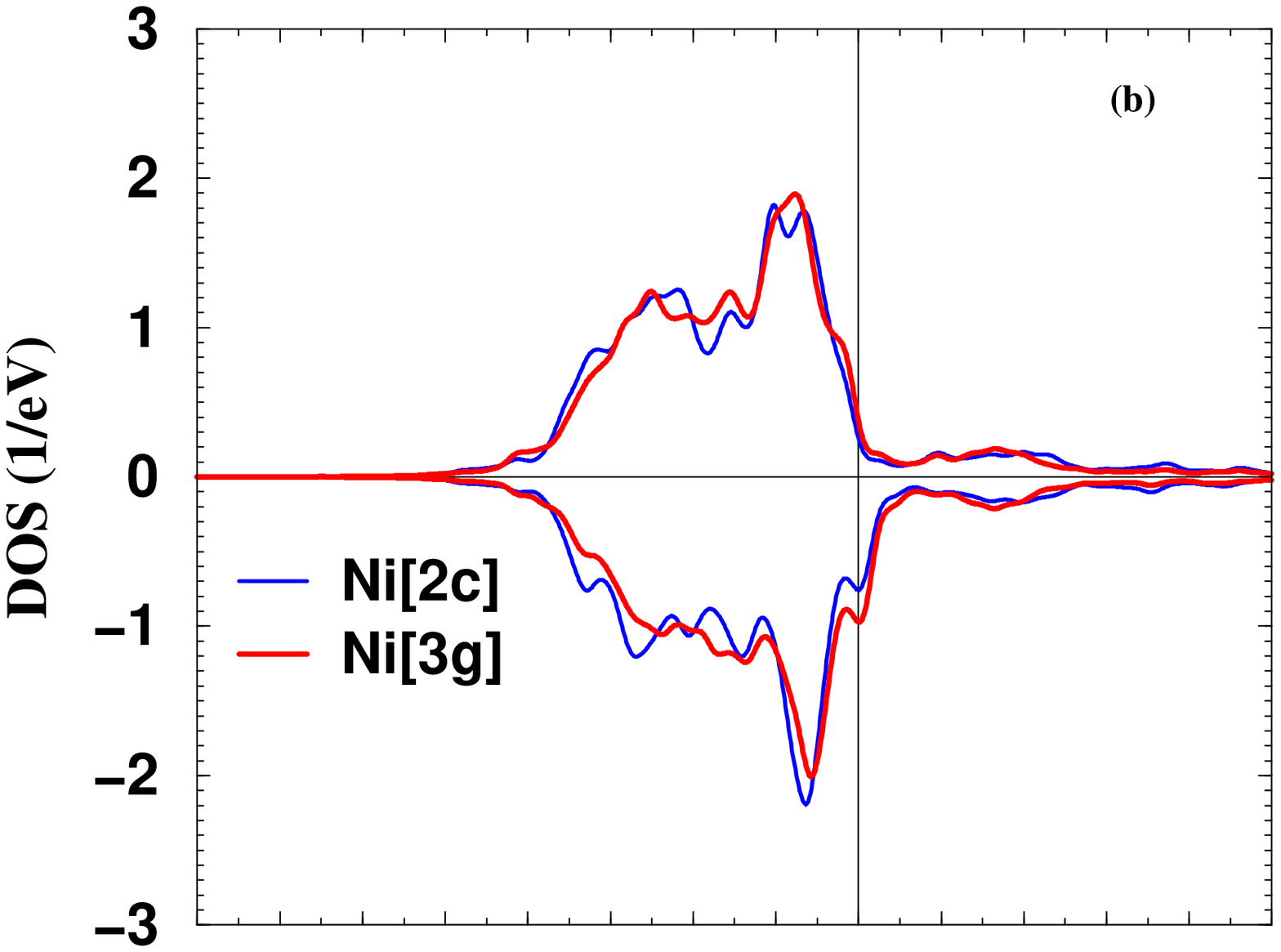}
\includegraphics[width=7cm]{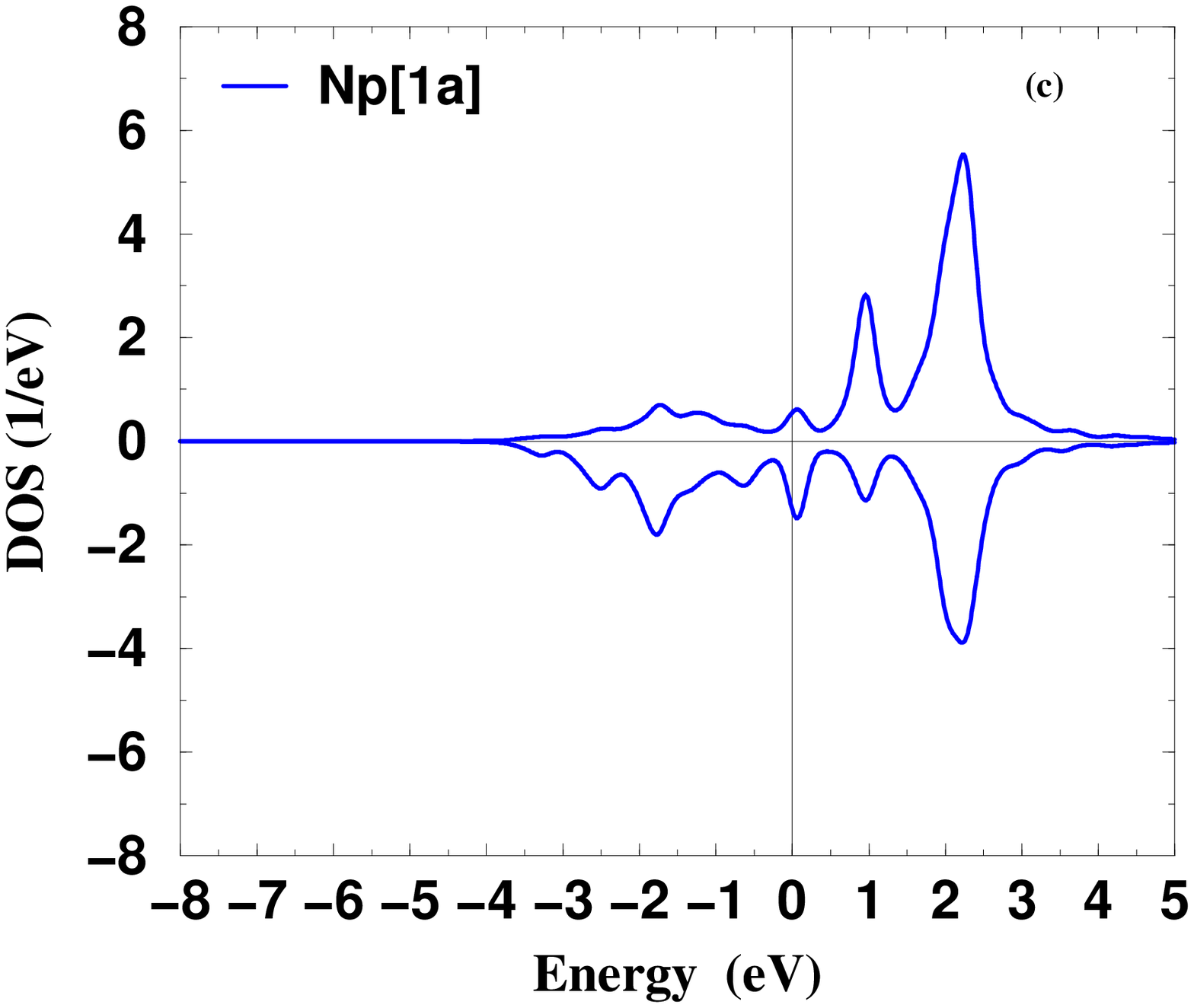}
\caption{(Color online) (a) The total electron-energy density of
states (DOS) per formula unit for NpNi$_{5}$  calculated with
relativistic LSDA+U ($U$=3 eV); (b) the $d$-orbital projected DOS
for Ni atoms at [2c], and [3g] special positions of the
$P6/mmm$ space group; (c) the $f$-orbital projected DOS for Np
atoms at [1a] position. \label{FigSashaNpNi5}}
\end{figure}

The total electron-energy density of states
(DOS) calculated for $U$ = 3~eV is shown in Fig.~\ref{FigSashaNpNi5}, together
with spin-resolved $d$-orbital projected DOS for Ni atoms at the
occupied lattice sites and $f$-orbital projected DOS for Np atom.
The Ni-$d^{\uparrow}$ band is
practically full, whereas the Ni-$d^{\downarrow}$ band is partially
occupied. The DOS peak in the vicinity of the Fermi level ($E_F$)
has mostly Ni-$d^{\downarrow}$ character.

The occupied part of
Np-$f$-manifold is mostly located in the energy range from 1 eV to 3
eV below $E_F$, and the 5$f$ electrons are therefore essentially localized.
Their contribution to the spin ($\mu_S^{5f}= -2 \langle
S_z^{5f}\rangle \mu_B/\hbar$) and orbital ($\mu_L^{5f} = -\langle
L_z^{5f}\rangle \mu_B/\hbar$) magnetic moments as well as the
magnetic dipole $\mu_{md}^{5f} = -6 \langle T_z \rangle \mu_B/\hbar$
contribution are given in Table ~\ref{occup}, together with the
calculated occupation numbers ($n^{5f}$, $n^{5f}_{5/2}$ and
$n^{5f}_{7/2}$), the Np $M_{4,5}$-edges branching ratio $B$, and the
valence spin-orbit interaction $\langle\vec{\ell}\cdot\vec{s}\rangle$ (from Eq.~\ref{so}).

These quantities can be directly compared with those deduced from
the shell-specific XMCD spectra measured at the Np $M_{4,5}$
absorption edges.

\begin{table}[htbp]
\caption{5$f$-states occupations $n^{5f}$, $n^{5f}_{5/2}$ and
$n^{5f}_{7/2}$, branching ratio $B$, and $5f$-electron contribution
to the valence spin-orbit interaction per hole  $w^{110}/n_{h}$, and
the spin, orbital, and magnetic-dipole moments ($\mu_S^{5f}$,
$\mu_L^{5f}$, and $\mu_{md}$) in Bohr magneton ($\mu_B$) units.}
\label{occup}
\begin{ruledtabular}
\begin{tabular}{ccccccccc}
 Atom[Site]     & $n^{5f}$  & $n^{5f}_{5/2}$ &  $n^{5f}_{7/2}$ & $B$ & $\langle \vec{\ell} \cdot \vec{s} \rangle$ &
 $\mu_S^{5f}$ & $\mu_L^{5f}$ & $\mu_{md}$\\
\hline
\multicolumn{2}{c}{LSDA(+SOC)}\\
\hline
Np$[1a]$ & 3.73 & 2.59 & 1.15 & 0.69 & -3.45 &-2.74 & 2.90 & -0.37 \\
\hline
\multicolumn{2}{c}{AMF-LSDA+U}\\
\hline
Np$[1a]$ & 3.77 & 3.24 & 0.53 & 0.75 & -5.68 &-1.63 & 3.98 & -2.81 \\
\end{tabular}
\end{ruledtabular}
\end{table}

The 5$f$ occupation number  is about 3.8, closer to the Np$^{3+} $ case.
Using the values from Tab.~\ref{occup} one obtains $\langle L_\zeta \rangle$ = -3.98,
$\langle S_\zeta \rangle + 3 \langle T_\zeta \rangle$ = 2.22, $\langle \vec{\ell} \cdot \vec{s} \rangle$=-5.68
and the branching ratio $B$=0.75, together with the $f$-shell magnetization of 2.3 $\mu_B$. All of these
numbers are in a reasonable agreement with the experimental data listed in Tab.~\ref{tab:Np3vsNp4XMCD} for
Np$^{3+}$.

\section{Discussion}
A comparison between XMCD results and first-principle calculations supports the assumption that the Np ions in NpNi$_{5}$ are trivalent. The Np ordered magnetic moment at 4 K, about 20$\%$ smaller than the Np$^{3+}$ free ion values  ($\mu$ = 2.57 $\mu_{B}$), suggests a rather localized nature of the 5$f$ electrons. This is supported by the results of first-principle calculations that locate the Np-$f$-manifold in the energy range from 1 eV to 3
eV below the Fermi level. The moment reduction can be attributed to the crystal-field potential. The $f$-electron behavior is therefore similar to the rare-earth analogs. However, whereas for the latter the Ni $d$ band is completely filled and magnetically inert, here one component of the spin-polarized Ni $d$ band is only partially filled. This is reflected in the value of the effective paramagnetic moment, which is higher than the free ion value for Np$^{3+}$.

The specific heat data suggest that the ground state is a doublet and a simple mean-field calculation is sufficient to obtain a good agreement with the values observed for the ordered moment and its spin and angular components. In the analog RNi$_{5}$ series, the sign of B$_{2}^{0}$ usually determines the preferred orientation of the magnetic moment, which is along the $c$-axis for negative B$_{2}^{0}$ (\textit{e.g.} for Er, Tm), and in the basal plane if B$_{2}^{0} >$ 0 (\textit{e.g.} for Nd, Tb, Dy, Ho) \cite{gignoux97} although, as observed for other rare-earth intermetallic systems \cite{caciuffo90,caciuffo95}, high-order terms in the crystal field potential are expected to play an important role in determining the magnetic anisotropy. For Np$^{3+}$, the Stevens factor $\alpha_{J}$ is positive ($\alpha_{J}$ = 9.13 $\times$ 10$^{-3}$ in intermediate coupling approximation \cite{amoretti84}), whereas the lattice electric field gradient $V^{latt}_{zz}$ should be positive, as observed for the isostructural GdNi$_{5}$ compound \cite{vansteenwijk77}. Thus B$_{2}^{0} \propto -\alpha_{J} V^{latt}_{zz}$ is negative in NpNi$_{5}$, the Np magnetic moments point along the $c$-axis, and the ground state is a $\Gamma_{5}$ doublet (note that in the ordered state of NpNi$_{5}$  the $z$ axis of the electric field gradient $V_{zz}$ is given by the direction of the ordered magnetic moment).

\begin{figure}
\includegraphics[width=7cm]{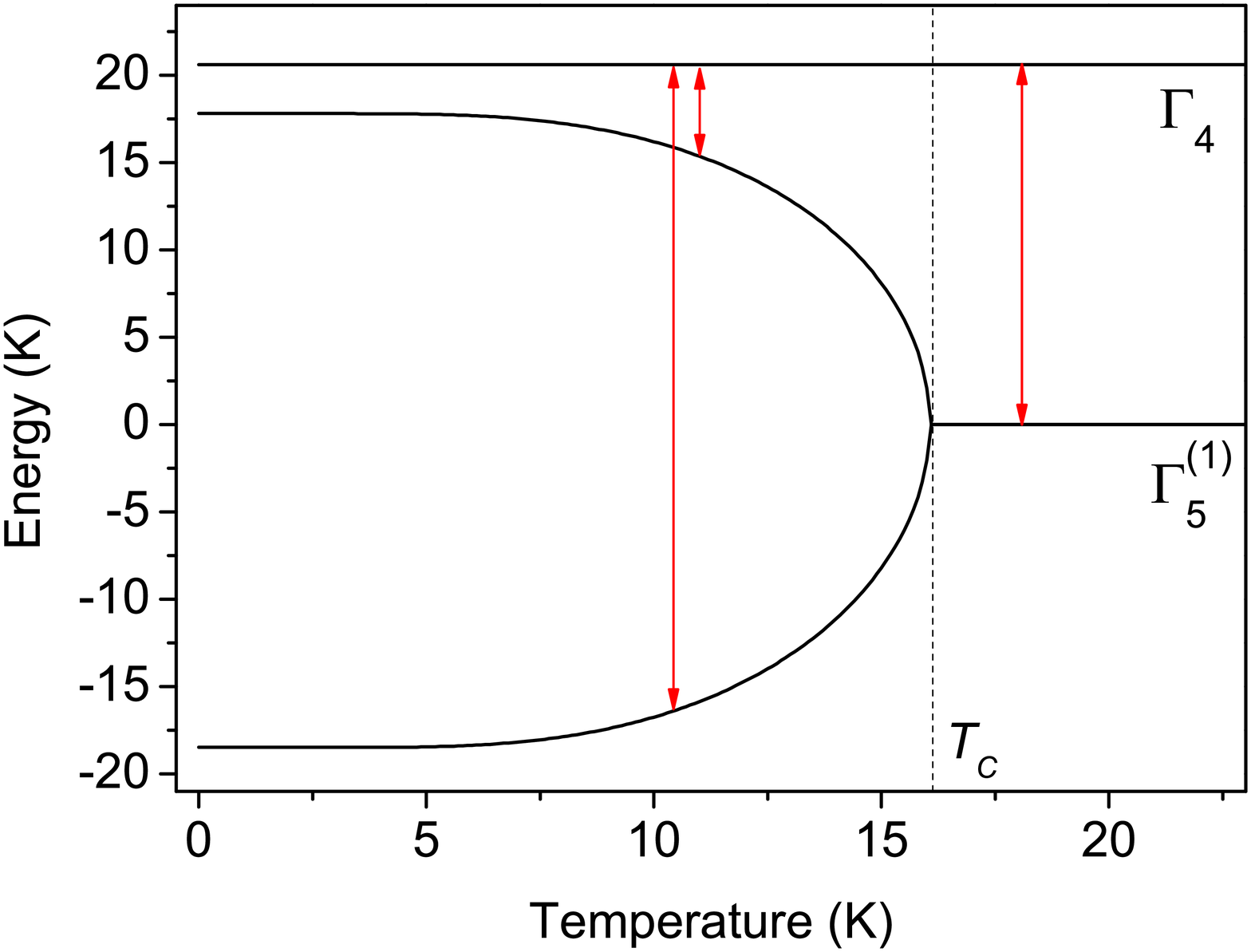}
\caption{Splitting of the low-energy crystal field levels by the molecular field in the ordered phase. The arrows indicate the allowed spin relaxation $\Delta J_{z} = \pm 1$ pathways between
the levels of the ground quasi-triplet. \label{levcross}}
\end{figure}

The electronic 5$f$ contribution to the quadrupole interaction is proportional to $<3J_{z}^{2} - J(J+1)>$ $\simeq$ 23. For the Np$^{3+}$ free ion, $<3J_{z}^{2} - J(J+1)> =$ 28, and the quadrupole interaction due to the 5$f$ electrons amounts to -27.3 mm/s \cite{sanchez88}. It follows that $eV_{zz}^{5f}Q(1-R)$ is negative and equal to -22.45 mm/s. In addition to this electronic contribution, the quadrupole interaction acting on the Np nuclei contains also a contribution from the lattice and the conduction electrons, $eV^{latt}_{zz}Q(1-\gamma_{\infty})$, where $\gamma_{\infty}$ is the Sternheimer factor. A rough estimate of this contribution can be provided by the quadrupole interaction measured for GdNi$_{5}$\cite{vansteenwijk77}, where the $f$ contribution to $V_{zz}$ vanishes, as Gd$^{3+}$ are $S$-state ions. From $eV_{zz}Q$ = 4.68 mm/s in GdNi$_{5}$ and using the quadrupole moment $Q$ = 1.3 $\times$ 10$^{-24}$ cm$^{2}$ for $^{155}$Gd nuclei \cite{tanaka92}, it follows that $V_{zz}^{latt} (1-\gamma_{\infty})$ = 10.4 $\times$ 10$^{-7}$ V/cm$^{2}$. Assuming that $V_{zz}^{latt}$ for both NpNi$_{5}$ and GdNi$_{5}$ are identical, and taking the values of $\gamma_{\infty}$ of -92 and -130 for Gd and Np, respectively \cite{barton79,thevenin87}, one obtains for NpNi$_{5}$ $V_{zz}^{latt} (1-\gamma_{\infty})$ = 14.6 $\times$ 10$^{17}$ V/cm$^{2}$ and $eV_{zz}^{latt} (1-\gamma_{\infty})Q$ = 30.2 mm/s, using the value of 4.1 $\times$ 10$^{-24}$ cm$^{2}$ for the quadrupole moment of the $^{237}$Np nuclei ground state \cite{dunlap85} (1 mm/s = 19.86$\times$10$^{-8}$ eV). As the principal axis of the lattice contribution to the electric field gradient is also parallel to the crystallographic $c$-axis, the estimated total quadrupole interaction in the low temperature ordered state is $eV_{zz}Q$ = 7.8 mm/s, relatively close to the experimental value of -10.5 mm/s. The difference between observed and calculated values can be attributed to the approximations made above, namely a fully localized 5$f$ electronic configuration, and equal lattice contributions for NpNi$_{5}$ and GdNi$_{5}$.

As shown in table \ref{tab:Np3vsNp4CF}, in the paramagnetic phase the low-energy crystal field level scheme is characterized by a doublet ground state with an excited singlet at 21 K. Direct spin-relaxation is highly suppressed within the $\Gamma_5$ doublet but allowed via the excited singlet \cite{kalvius85,gal85}. As long as the relaxation rate is slow, we observe the coexistence of a magnetic split pattern and of a somewhat broadened quadrupole spectrum, for example at 20 K (Fig. \ref{MossNpNi5}). With increasing temperature, the singlet becomes more and more populated, the spin-relaxation rate speeds up, the magnetic split pattern collapses and the quadrupolar spectrum becomes less and less broad, until the limit of fast relaxation is reached. Below T$_C$, the doublet is split by the molecular field, as depicted in Fig. \ref{levcross}. However, a quadrupole split spectrum is still observed, together with the full-split pattern originating from the ground state, as long as the singlet is thermally populated. When the temperature is low enough, only the ground state is populated, the relaxation slows down and the hyperfine field is almost temperature independent.

\section{Conclusions}
The low temperature physical behavior of NpNi$_{5}$ has been investigated by magnetization, M\"{o}ssbauer spectroscopy, specific heat, and x-ray magnetic circular dichroism measurements. The compound crystallizes in the CaCu$_{5}$-type hexagonal structure and, at T$_{C}$ $\sim$ 16 K, undergoes a transition to a ferromagnetic state. The results suggest a trivalent oxidation state for neptunium (5\textit{f}$^{4}$ electronic configuration) and an ordered moment at the Np site of $\sim$ 2 $\mu_{B}$ is obtained by  M\"{o}ssbauer spectroscopy. The Ni up-spin 3\textit{d} band is almost filled whereas the down-spin one is only partially occupied, resulting in a magnetic moment at the Ni sites of about 0.08 $\mu_{B}$.
The peculiar temperature dependence of the M\"{o}ssbauer spectra is explained by the presence of two separate relaxation phenomena within the sublevels of the ground quasi-triplet.

Combining XMCD and M\"{o}ssbauer spectroscopy results, we obtain the spin and orbital contributions to the Np magnetic moment
($\mu_{S}$ = -1.88~$\mu_{B}$ and $\mu_{L}$ = 3.91~$\mu_{B}$). The ratio between the expectation value of the magnetic dipole operator and the spin magnetic moment is positive and large, as predicted for localized 5$f$ electrons. The expectation value of the angular part of the spin-orbit interaction operator is in good agreement with the value calculated in intermediate coupling approximation for Np ions.
The temperature dependence of the specific heat shows a narrow anomaly at T$_{C}$, which broadens and shifts towards higher temperatures when an external magnetic field is applied. The vibrational contribution to the specific heat has been estimated from the curve measured for ThNi$_5$. This allowed us to determine the magnetic entropy change that gives information on the electronic energy level distribution.  Although the 5$f$-electron narrow band is well below the Fermi energy, the overall magnetic behavior is essentially different form the localized rare earth case, mostly due to the contributions of the nickel sublattice.

\acknowledgments
{We thank D. Bou\"{e}xi\`{e}re, G. Pagliosa, and P. Amador, for their technical support. The high-purity neptunium metal required for the fabrication of the sample was made available through a loan agreement between Lawrence Livermore National Laboratory and ITU, in the frame of a collaboration involving LLNL, Los Alamos National Laboratory, and the U.S. Department of Energy. A. H. acknowledges the European Commission for support in the frame of the Training and Mobility of Researchers programme. The support from Czech Republic grant GACR P204/10/0330 is acknowledged.
}

\bibliography{NpNi5}

\begin{thebibliography}{54}
\expandafter\ifx\csname natexlab\endcsname\relax\def\natexlab#1{#1}\fi
\expandafter\ifx\csname bibnamefont\endcsname\relax
  \def\bibnamefont#1{#1}\fi
\expandafter\ifx\csname bibfnamefont\endcsname\relax
  \def\bibfnamefont#1{#1}\fi
\expandafter\ifx\csname citenamefont\endcsname\relax
  \def\citenamefont#1{#1}\fi
\expandafter\ifx\csname url\endcsname\relax
  \def\url#1{\texttt{#1}}\fi
\expandafter\ifx\csname urlprefix\endcsname\relax\def\urlprefix{URL }\fi
\providecommand{\bibinfo}[2]{#2}
\providecommand{\eprint}[2][]{\url{#2}}

\bibitem[{\citenamefont{Akabori et~al.}(1997)\citenamefont{Akabori, Haire,
  Gibson, Okamoto, and Ogawa}}]{akabori97}
\bibinfo{author}{\bibfnamefont{M.}~\bibnamefont{Akabori}},
  \bibinfo{author}{\bibfnamefont{R.~G.} \bibnamefont{Haire}},
  \bibinfo{author}{\bibfnamefont{J.~K.} \bibnamefont{Gibson}},
  \bibinfo{author}{\bibfnamefont{Y.}~\bibnamefont{Okamoto}}, \bibnamefont{and}
  \bibinfo{author}{\bibfnamefont{T.}~\bibnamefont{Ogawa}}, \bibinfo{journal}{J.
  Alloys \& Compnds.} \textbf{\bibinfo{volume}{257}}, \bibinfo{pages}{268 }
  (\bibinfo{year}{1997}).

\bibitem[{\citenamefont{Zachariasen}(1973)}]{zachariasen73}
\bibinfo{author}{\bibfnamefont{W.~H.} \bibnamefont{Zachariasen}},
  \bibinfo{journal}{J. Inorg. Nucl. Chem.} \textbf{\bibinfo{volume}{35}},
  \bibinfo{pages}{3487} (\bibinfo{year}{1973}).

\bibitem[{\citenamefont{Hill}(1971)}]{hill71}
\bibinfo{author}{\bibfnamefont{H.}~\bibnamefont{Hill}},
  \bibinfo{journal}{\textit{Plutonium 1970 and Other Actinides}, W. N. Miner,
  Ed.,} pp. \bibinfo{pages}{2, AIME, New York} (\bibinfo{year}{1971}).

\bibitem[{\citenamefont{van Vucht et~al.}(1970)\citenamefont{van Vucht,
  Kuijpers, and Bruning}}]{vanvucht70}
\bibinfo{author}{\bibfnamefont{J.~H.~N.} \bibnamefont{van Vucht}},
  \bibinfo{author}{\bibfnamefont{F.~A.} \bibnamefont{Kuijpers}},
  \bibnamefont{and} \bibinfo{author}{\bibfnamefont{H.~C. A.~M.}
  \bibnamefont{Bruning}}, \bibinfo{journal}{Philips Res. Rep.}
  \textbf{\bibinfo{volume}{25}}, \bibinfo{pages}{1335} (\bibinfo{year}{1970}).

\bibitem[{\citenamefont{Kuijpers}(1973)}]{kuijpers73}
\bibinfo{author}{\bibfnamefont{F.~A.} \bibnamefont{Kuijpers}},
  \bibinfo{journal}{Philips Res. Rep.} \textbf{\bibinfo{volume}{2}},
  \bibinfo{pages}{1} (\bibinfo{year}{1973}).

\bibitem[{\citenamefont{Takeshita et~al.}(1980)\citenamefont{Takeshita,
  Gschneider~Jr., Thome, and McMasters}}]{takeshita80}
\bibinfo{author}{\bibfnamefont{T.}~\bibnamefont{Takeshita}},
  \bibinfo{author}{\bibfnamefont{K.~A.} \bibnamefont{Gschneider~Jr.}},
  \bibinfo{author}{\bibfnamefont{D.~K.} \bibnamefont{Thome}}, \bibnamefont{and}
  \bibinfo{author}{\bibfnamefont{O.~D.} \bibnamefont{McMasters}},
  \bibinfo{journal}{Phys. Rev. B} \textbf{\bibinfo{volume}{21}},
  \bibinfo{pages}{5636} (\bibinfo{year}{1980}).

\bibitem[{\citenamefont{Szajek et~al.}(2003)\citenamefont{Szajek, Jurczyk,
  Nowak, and Makowiecka}}]{szajek03}
\bibinfo{author}{\bibfnamefont{A.}~\bibnamefont{Szajek}},
  \bibinfo{author}{\bibfnamefont{M.}~\bibnamefont{Jurczyk}},
  \bibinfo{author}{\bibfnamefont{M.}~\bibnamefont{Nowak}}, \bibnamefont{and}
  \bibinfo{author}{\bibfnamefont{M.}~\bibnamefont{Makowiecka}},
  \bibinfo{journal}{Phys. Status Solidi (a)} \textbf{\bibinfo{volume}{196}},
  \bibinfo{pages}{252} (\bibinfo{year}{2003}).

\bibitem[{\citenamefont{Yu et~al.}(2008)\citenamefont{Yu, Han, Zhao, Xue, and
  Gao}}]{yu08}
\bibinfo{author}{\bibfnamefont{Y.}~\bibnamefont{Yu}},
  \bibinfo{author}{\bibfnamefont{H.}~\bibnamefont{Han}},
  \bibinfo{author}{\bibfnamefont{Y.}~\bibnamefont{Zhao}},
  \bibinfo{author}{\bibfnamefont{W.}~\bibnamefont{Xue}}, \bibnamefont{and}
  \bibinfo{author}{\bibfnamefont{T.}~\bibnamefont{Gao}},
  \bibinfo{journal}{Solid State Commun.} \textbf{\bibinfo{volume}{148}},
  \bibinfo{pages}{1 } (\bibinfo{year}{2008}).

\bibitem[{\citenamefont{Goremychkin et~al.}(1985)\citenamefont{Goremychkin,
  M\"{u}hle, Lippold, Chistyakov, and Savitskii}}]{goremychkin85}
\bibinfo{author}{\bibfnamefont{E.~A.} \bibnamefont{Goremychkin}},
  \bibinfo{author}{\bibfnamefont{E.}~\bibnamefont{M\"{u}hle}},
  \bibinfo{author}{\bibfnamefont{B.}~\bibnamefont{Lippold}},
  \bibinfo{author}{\bibfnamefont{O.~D.} \bibnamefont{Chistyakov}},
  \bibnamefont{and} \bibinfo{author}{\bibfnamefont{E.~M.}
  \bibnamefont{Savitskii}}, \bibinfo{journal}{Phys. Status Solidi (b)}
  \textbf{\bibinfo{volume}{127}}, \bibinfo{pages}{371} (\bibinfo{year}{1985}).

\bibitem[{\citenamefont{Barthem et~al.}(1988)\citenamefont{Barthem, Gignoux,
  Na\"{\i}t-Saada, Schmitt, and Creuzet}}]{barthem88}
\bibinfo{author}{\bibfnamefont{V.~M. T.~S.} \bibnamefont{Barthem}},
  \bibinfo{author}{\bibfnamefont{D.}~\bibnamefont{Gignoux}},
  \bibinfo{author}{\bibfnamefont{A.}~\bibnamefont{Na\"{\i}t-Saada}},
  \bibinfo{author}{\bibfnamefont{D.}~\bibnamefont{Schmitt}}, \bibnamefont{and}
  \bibinfo{author}{\bibfnamefont{G.}~\bibnamefont{Creuzet}},
  \bibinfo{journal}{Phys. Rev. B} \textbf{\bibinfo{volume}{37}},
  \bibinfo{pages}{1733} (\bibinfo{year}{1988}).

\bibitem[{\citenamefont{Reiffers et~al.}(1989)\citenamefont{Reiffers, Naidyuk,
  Jansen, Wyder, Yanson, Gignoux, and Schmitt}}]{reiffers89}
\bibinfo{author}{\bibfnamefont{M.}~\bibnamefont{Reiffers}},
  \bibinfo{author}{\bibfnamefont{Y.~G.} \bibnamefont{Naidyuk}},
  \bibinfo{author}{\bibfnamefont{A.~G.~M.} \bibnamefont{Jansen}},
  \bibinfo{author}{\bibfnamefont{P.}~\bibnamefont{Wyder}},
  \bibinfo{author}{\bibfnamefont{I.~K.} \bibnamefont{Yanson}},
  \bibinfo{author}{\bibfnamefont{D.}~\bibnamefont{Gignoux}}, \bibnamefont{and}
  \bibinfo{author}{\bibfnamefont{D.~P.} \bibnamefont{Schmitt}},
  \bibinfo{journal}{Phys. Rev. Lett.} \textbf{\bibinfo{volume}{62}},
  \bibinfo{pages}{1560} (\bibinfo{year}{1989}).

\bibitem[{\citenamefont{Gubbens et~al.}(1989)\citenamefont{Gubbens, van~der
  Kraan, and Buschow}}]{gubbens89}
\bibinfo{author}{\bibfnamefont{P.~C.~M.} \bibnamefont{Gubbens}},
  \bibinfo{author}{\bibfnamefont{A.~M.} \bibnamefont{van~der Kraan}},
  \bibnamefont{and} \bibinfo{author}{\bibfnamefont{K.~H.~J.}
  \bibnamefont{Buschow}}, \bibinfo{journal}{Phys. Rev. B}
  \textbf{\bibinfo{volume}{39}}, \bibinfo{pages}{12548} (\bibinfo{year}{1989}).

\bibitem[{\citenamefont{Amato et~al.}(1992)\citenamefont{Amato, B\"{u}hrer,
  Grayevsky, Gygax, Furrer, Kaplan, and Schenck}}]{amato92}
\bibinfo{author}{\bibfnamefont{A.}~\bibnamefont{Amato}},
  \bibinfo{author}{\bibfnamefont{W.}~\bibnamefont{B\"{u}hrer}},
  \bibinfo{author}{\bibfnamefont{A.}~\bibnamefont{Grayevsky}},
  \bibinfo{author}{\bibfnamefont{F.~N.} \bibnamefont{Gygax}},
  \bibinfo{author}{\bibfnamefont{A.}~\bibnamefont{Furrer}},
  \bibinfo{author}{\bibfnamefont{N.}~\bibnamefont{Kaplan}}, \bibnamefont{and}
  \bibinfo{author}{\bibfnamefont{A.}~\bibnamefont{Schenck}},
  \bibinfo{journal}{Solid State Commun.} \textbf{\bibinfo{volume}{82}},
  \bibinfo{pages}{767 } (\bibinfo{year}{1992}).

\bibitem[{\citenamefont{Nov\'{a}k and Kuriplach}(1994)}]{novak94}
\bibinfo{author}{\bibfnamefont{P.}~\bibnamefont{Nov\'{a}k}} \bibnamefont{and}
  \bibinfo{author}{\bibfnamefont{J.}~\bibnamefont{Kuriplach}},
  \bibinfo{journal}{Phys. Rev. B} \textbf{\bibinfo{volume}{50}},
  \bibinfo{pages}{2085} (\bibinfo{year}{1994}).

\bibitem[{\citenamefont{Svoboda et~al.}(2004)\citenamefont{Svoboda,
  Vejpravov\'{a}, Kim-Ngan, and Kaysel}}]{svoboda04}
\bibinfo{author}{\bibfnamefont{P.}~\bibnamefont{Svoboda}},
  \bibinfo{author}{\bibfnamefont{J.}~\bibnamefont{Vejpravov\'{a}}},
  \bibinfo{author}{\bibfnamefont{N.-T.} \bibnamefont{Kim-Ngan}},
  \bibnamefont{and} \bibinfo{author}{\bibfnamefont{F.}~\bibnamefont{Kaysel}},
  \bibinfo{journal}{J. Magn. Magn. Mater.} \textbf{\bibinfo{volume}{272-276}},
  \bibinfo{pages}{595 } (\bibinfo{year}{2004}).

\bibitem[{\citenamefont{Buschow}(1977)}]{buschow77}
\bibinfo{author}{\bibfnamefont{K.~H.~J.} \bibnamefont{Buschow}},
  \bibinfo{journal}{Rep. Prog. Phys.} \textbf{\bibinfo{volume}{40}},
  \bibinfo{pages}{1179} (\bibinfo{year}{1977}).

\bibitem[{\citenamefont{Gignoux et~al.}(1976)\citenamefont{Gignoux, Givord, and
  Del~Moral}}]{gignoux76}
\bibinfo{author}{\bibfnamefont{D.}~\bibnamefont{Gignoux}},
  \bibinfo{author}{\bibfnamefont{F.}~\bibnamefont{Givord}}, \bibnamefont{and}
  \bibinfo{author}{\bibfnamefont{A.}~\bibnamefont{Del~Moral}},
  \bibinfo{journal}{Solid State Commun.} \textbf{\bibinfo{volume}{19}},
  \bibinfo{pages}{89} (\bibinfo{year}{1976}).

\bibitem[{\citenamefont{Folle et~al.}(1981)\citenamefont{Folle, Kubota, Buchal,
  Mueller, and Pobell}}]{folle81}
\bibinfo{author}{\bibfnamefont{H.}~\bibnamefont{Folle}},
  \bibinfo{author}{\bibfnamefont{M.}~\bibnamefont{Kubota}},
  \bibinfo{author}{\bibfnamefont{C.}~\bibnamefont{Buchal}},
  \bibinfo{author}{\bibfnamefont{R.}~\bibnamefont{Mueller}}, \bibnamefont{and}
  \bibinfo{author}{\bibfnamefont{F.}~\bibnamefont{Pobell}},
  \bibinfo{journal}{Z. Phys. B}
  \textbf{\bibinfo{volume}{41}}, \bibinfo{pages}{223} (\bibinfo{year}{1981}).

\bibitem[{\citenamefont{Andres and Darack}(1977)}]{andres77}
\bibinfo{author}{\bibfnamefont{K.}~\bibnamefont{Andres}} \bibnamefont{and}
  \bibinfo{author}{\bibfnamefont{S.}~\bibnamefont{Darack}},
  \bibinfo{journal}{Physica B+C} \textbf{\bibinfo{volume}{86-88}},
  \bibinfo{pages}{1071} (\bibinfo{year}{1977}).

\bibitem[{\citenamefont{Kuchin et~al.}(2004)\citenamefont{Kuchin, Gurevich,
  Dmitriev, Terekhov, Chagovets, and Ermolenko}}]{kuchin04}
\bibinfo{author}{\bibfnamefont{A.~G.} \bibnamefont{Kuchin}},
  \bibinfo{author}{\bibfnamefont{A.~M.} \bibnamefont{Gurevich}},
  \bibinfo{author}{\bibfnamefont{V.~M.} \bibnamefont{Dmitriev}},
  \bibinfo{author}{\bibfnamefont{A.~V.} \bibnamefont{Terekhov}},
  \bibinfo{author}{\bibfnamefont{T.~V.} \bibnamefont{Chagovets}},
  \bibnamefont{and} \bibinfo{author}{\bibfnamefont{A.~S.}
  \bibnamefont{Ermolenko}}, \bibinfo{journal}{J. Alloys \& Compnds.}
  \textbf{\bibinfo{volume}{368}}, \bibinfo{pages}{75} (\bibinfo{year}{2004}).

\bibitem[{\citenamefont{Kuchin et~al.}(2006)\citenamefont{Kuchin, Ermolenko,
  Kulikov, Khrabrov, Rosenfeld, Makarova, Lapina, and Belozerov}}]{kuchin06}
\bibinfo{author}{\bibfnamefont{A.~G.} \bibnamefont{Kuchin}},
  \bibinfo{author}{\bibfnamefont{A.~S.} \bibnamefont{Ermolenko}},
  \bibinfo{author}{\bibfnamefont{Y.~A.} \bibnamefont{Kulikov}},
  \bibinfo{author}{\bibfnamefont{V.~I.} \bibnamefont{Khrabrov}},
  \bibinfo{author}{\bibfnamefont{E.~V.} \bibnamefont{Rosenfeld}},
  \bibinfo{author}{\bibfnamefont{G.~M.} \bibnamefont{Makarova}},
  \bibinfo{author}{\bibfnamefont{T.~P.} \bibnamefont{Lapina}},
  \bibnamefont{and} \bibinfo{author}{\bibfnamefont{Y.~V.}
  \bibnamefont{Belozerov}}, \bibinfo{journal}{J. Magn. Magn. Mater.}
  \textbf{\bibinfo{volume}{303}}, \bibinfo{pages}{119} (\bibinfo{year}{2006}).

\bibitem[{\citenamefont{Santini et~al.}(2006)\citenamefont{Santini, Carretta,
  Magnani, Amoretti, and Caciuffo}}]{santini06}
\bibinfo{author}{\bibfnamefont{P.}~\bibnamefont{Santini}},
  \bibinfo{author}{\bibfnamefont{S.}~\bibnamefont{Carretta}},
  \bibinfo{author}{\bibfnamefont{N.}~\bibnamefont{Magnani}},
  \bibinfo{author}{\bibfnamefont{G.}~\bibnamefont{Amoretti}}, \bibnamefont{and}
  \bibinfo{author}{\bibfnamefont{R.}~\bibnamefont{Caciuffo}},
  \bibinfo{journal}{Phys. Rev. Lett.} \textbf{\bibinfo{volume}{97}},
  \bibinfo{pages}{207203} (\bibinfo{year}{2006}).

\bibitem[{\citenamefont{Magnani et~al.}(2008)\citenamefont{Magnani, Carretta,
  Caciuffo, Santini, Amoretti, Hiess, Rebizant, and Lander}}]{magnani08}
\bibinfo{author}{\bibfnamefont{N.}~\bibnamefont{Magnani}},
  \bibinfo{author}{\bibfnamefont{S.}~\bibnamefont{Carretta}},
  \bibinfo{author}{\bibfnamefont{R.}~\bibnamefont{Caciuffo}},
  \bibinfo{author}{\bibfnamefont{P.}~\bibnamefont{Santini}},
  \bibinfo{author}{\bibfnamefont{G.}~\bibnamefont{Amoretti}},
  \bibinfo{author}{\bibfnamefont{A.}~\bibnamefont{Hiess}},
  \bibinfo{author}{\bibfnamefont{J.}~\bibnamefont{Rebizant}}, \bibnamefont{and}
  \bibinfo{author}{\bibfnamefont{G.~H.} \bibnamefont{Lander}},
  \bibinfo{journal}{Phys. Rev. B} \textbf{\bibinfo{volume}{78}},
  \bibinfo{pages}{104425} (\bibinfo{year}{2008}).

\bibitem[{\citenamefont{Barthem et~al.}(1989)\citenamefont{Barthem, Gignoux,
  Nait-Saada, Schmitt, and Takeuchi}}]{barthem89}
\bibinfo{author}{\bibfnamefont{V.~M. T.~S.} \bibnamefont{Barthem}},
  \bibinfo{author}{\bibfnamefont{D.}~\bibnamefont{Gignoux}},
  \bibinfo{author}{\bibfnamefont{A.}~\bibnamefont{Nait-Saada}},
  \bibinfo{author}{\bibfnamefont{D.}~\bibnamefont{Schmitt}}, \bibnamefont{and}
  \bibinfo{author}{\bibfnamefont{A.~Y.} \bibnamefont{Takeuchi}},
  \bibinfo{journal}{J. Magn. Magn. Mater.} \textbf{\bibinfo{volume}{80}},
  \bibinfo{pages}{142 } (\bibinfo{year}{1989}).

\bibitem[{\citenamefont{Halevy et~al.}(2012)\citenamefont{Halevy, Hen, Orion,
  Colineau, Eloirdi, Griveau, Gaczy\ifmmode~\acute{n}\else \'{n}\fi{}ski,
  Wilhelm, Rogalev, Sanchez et~al.}}]{halevy12}
\bibinfo{author}{\bibfnamefont{I.}~\bibnamefont{Halevy}},
  \bibinfo{author}{\bibfnamefont{A.}~\bibnamefont{Hen}},
  \bibinfo{author}{\bibfnamefont{I.}~\bibnamefont{Orion}},
  \bibinfo{author}{\bibfnamefont{E.}~\bibnamefont{Colineau}},
  \bibinfo{author}{\bibfnamefont{R.}~\bibnamefont{Eloirdi}},
  \bibinfo{author}{\bibfnamefont{J.-C.} \bibnamefont{Griveau}},
  \bibinfo{author}{\bibfnamefont{P.}~\bibnamefont{Gaczy\ifmmode~\acute{n}\else
  \'{n}\fi{}ski}}, \bibinfo{author}{\bibfnamefont{F.}~\bibnamefont{Wilhelm}},
  \bibinfo{author}{\bibfnamefont{A.}~\bibnamefont{Rogalev}},
  \bibinfo{author}{\bibfnamefont{J.-P.} \bibnamefont{Sanchez}},
  \bibinfo{author}{\bibfnamefont{M. L.} \bibnamefont{Winterrose}},
  \bibinfo{author}{\bibfnamefont{N.} \bibnamefont{Magnani}},
  \bibinfo{author}{\bibfnamefont{A. B.} \bibnamefont{Shick}}, \bibnamefont{and}
  \bibinfo{author}{\bibfnamefont{R.} \bibnamefont{Caciuffo}},
  \bibinfo{journal}{Phys. Rev. B}
  \textbf{\bibinfo{volume}{85}}, \bibinfo{pages}{014434}
  (\bibinfo{year}{2012}).

\bibitem[{\citenamefont{Dunlap and Kalvius}(1985)}]{dunlap85}
\bibinfo{author}{\bibfnamefont{B.~D.} \bibnamefont{Dunlap}} \bibnamefont{and}
  \bibinfo{author}{\bibfnamefont{G.~M.} \bibnamefont{Kalvius}}, in
  \emph{\bibinfo{booktitle}{Handbook on the Physics and Chemistry of the
  Actinides}}, edited by \bibinfo{editor}{\bibfnamefont{A.~J.}
  \bibnamefont{Freeman}} \bibnamefont{and}
  \bibinfo{editor}{\bibfnamefont{G.~H.} \bibnamefont{Lander}}
  (\bibinfo{publisher}{North-Holland, Amsterdam}, \bibinfo{year}{1985}),
  vol.~\bibinfo{volume}{2}, p. \bibinfo{pages}{329}.

\bibitem[{\citenamefont{Andres et~al.}(1979)\citenamefont{Andres, Darack, and
  Ott}}]{andres79}
\bibinfo{author}{\bibfnamefont{K.}~\bibnamefont{Andres}},
  \bibinfo{author}{\bibfnamefont{S.}~\bibnamefont{Darack}}, \bibnamefont{and}
  \bibinfo{author}{\bibfnamefont{H.~R.} \bibnamefont{Ott}},
  \bibinfo{journal}{Phys. Rev. B} \textbf{\bibinfo{volume}{19}},
  \bibinfo{pages}{5475} (\bibinfo{year}{1979}).

\bibitem[{\citenamefont{van~der Laan}(2013)}]{vanderlaan13}
\bibinfo{author}{\bibfnamefont{G.}~\bibnamefont{van~der Laan}},
  \bibinfo{journal}{J. Phys.: Conf. Ser.} \textbf{\bibinfo{volume}{430}},
  \bibinfo{pages}{012127} (\bibinfo{year}{2013}).

\bibitem[{\citenamefont{Goulon et~al.}(1982)\citenamefont{Goulon, Goulon-Ginet,
  Cortes, and Dubois}}]{goulon82}
\bibinfo{author}{\bibfnamefont{J.}~\bibnamefont{Goulon}},
  \bibinfo{author}{\bibfnamefont{C.}~\bibnamefont{Goulon-Ginet}},
  \bibinfo{author}{\bibfnamefont{R.}~\bibnamefont{Cortes}}, \bibnamefont{and}
  \bibinfo{author}{\bibfnamefont{J.~M.} \bibnamefont{Dubois}},
  \bibinfo{journal}{J. Phys. (Paris)} \textbf{\bibinfo{volume}{43}},
  \bibinfo{pages}{539} (\bibinfo{year}{1982}).

\bibitem[{\citenamefont{Tr\"{o}ger et~al.}(1992)\citenamefont{Tr\"{o}ger,
  Arvanitis, Baberschke, Michaelis, Grimm, and Zschech}}]{troger92}
\bibinfo{author}{\bibfnamefont{L.}~\bibnamefont{Tr\"{o}ger}},
  \bibinfo{author}{\bibfnamefont{D.}~\bibnamefont{Arvanitis}},
  \bibinfo{author}{\bibfnamefont{K.}~\bibnamefont{Baberschke}},
  \bibinfo{author}{\bibfnamefont{H.}~\bibnamefont{Michaelis}},
  \bibinfo{author}{\bibfnamefont{U.}~\bibnamefont{Grimm}}, \bibnamefont{and}
  \bibinfo{author}{\bibfnamefont{E.}~\bibnamefont{Zschech}},
  \bibinfo{journal}{Phys. Rev. B} \textbf{\bibinfo{volume}{46}},
  \bibinfo{pages}{3283} (\bibinfo{year}{1992}).

\bibitem[{\citenamefont{Pfalzer et~al.}(1999)\citenamefont{Pfalzer, Urbach,
  Klemm, Horn, denBoer, Frenkel, and Kirkland}}]{pfalzer99}
\bibinfo{author}{\bibfnamefont{P.}~\bibnamefont{Pfalzer}},
  \bibinfo{author}{\bibfnamefont{J.-P.} \bibnamefont{Urbach}},
  \bibinfo{author}{\bibfnamefont{M.}~\bibnamefont{Klemm}},
  \bibinfo{author}{\bibfnamefont{S.}~\bibnamefont{Horn}},
  \bibinfo{author}{\bibfnamefont{M.~L.} \bibnamefont{denBoer}},
  \bibinfo{author}{\bibfnamefont{A.~I.} \bibnamefont{Frenkel}},
  \bibnamefont{and} \bibinfo{author}{\bibfnamefont{J.~P.}
  \bibnamefont{Kirkland}}, \bibinfo{journal}{Phys. Rev. B}
  \textbf{\bibinfo{volume}{60}}, \bibinfo{pages}{9335} (\bibinfo{year}{1999}).

\bibitem[{\citenamefont{Berger et~al.}()\citenamefont{Berger, Hubbell, Seltzer,
  Chang, Coursey, Sukumar, Zucker, and Olsen}}]{berger}
\bibinfo{author}{\bibfnamefont{M.~J.} \bibnamefont{Berger}},
  \bibinfo{author}{\bibfnamefont{J.~H.} \bibnamefont{Hubbell}},
  \bibinfo{author}{\bibfnamefont{S.~M.} \bibnamefont{Seltzer}},
  \bibinfo{author}{\bibfnamefont{J.}~\bibnamefont{Chang}},
  \bibinfo{author}{\bibfnamefont{J.~S.} \bibnamefont{Coursey}},
  \bibinfo{author}{\bibfnamefont{R.}~\bibnamefont{Sukumar}},
  \bibinfo{author}{\bibfnamefont{D.~S.} \bibnamefont{Zucker}},
  \bibnamefont{and} \bibinfo{author}{\bibfnamefont{K.}~\bibnamefont{Olsen}},
  \urlprefix\url{http://www.nist.gov/pml/data/xcom}.

\bibitem[{\citenamefont{van~der Laan and Thole}(1996)}]{vanderlaan96}
\bibinfo{author}{\bibfnamefont{G.}~\bibnamefont{van~der Laan}}
  \bibnamefont{and} \bibinfo{author}{\bibfnamefont{B.~T.} \bibnamefont{Thole}},
  \bibinfo{journal}{Phys. Rev. B} \textbf{\bibinfo{volume}{53}},
  \bibinfo{pages}{14458} (\bibinfo{year}{1996}).

\bibitem[{\citenamefont{van~der Laan et~al.}(2004)\citenamefont{van~der Laan,
  Moore, Tobin, Chung, Wall, and Schwartz}}]{vanderlaan04}
\bibinfo{author}{\bibfnamefont{G.}~\bibnamefont{van~der Laan}},
  \bibinfo{author}{\bibfnamefont{K.~T.} \bibnamefont{Moore}},
  \bibinfo{author}{\bibfnamefont{J.~G.} \bibnamefont{Tobin}},
  \bibinfo{author}{\bibfnamefont{B.~W.} \bibnamefont{Chung}},
  \bibinfo{author}{\bibfnamefont{M.~A.} \bibnamefont{Wall}}, \bibnamefont{and}
  \bibinfo{author}{\bibfnamefont{A.~J.} \bibnamefont{Schwartz}},
  \bibinfo{journal}{Phys. Rev. Lett.} \textbf{\bibinfo{volume}{93}},
  \bibinfo{pages}{097401} (\bibinfo{year}{2004}).

\bibitem[{\citenamefont{Thole et~al.}(1992)\citenamefont{Thole, Carra, Sette,
  and van~der Laan}}]{thole92}
\bibinfo{author}{\bibfnamefont{B.~T.} \bibnamefont{Thole}},
  \bibinfo{author}{\bibfnamefont{P.}~\bibnamefont{Carra}},
  \bibinfo{author}{\bibfnamefont{F.}~\bibnamefont{Sette}}, \bibnamefont{and}
  \bibinfo{author}{\bibfnamefont{G.}~\bibnamefont{van~der Laan}},
  \bibinfo{journal}{Phys. Rev. Lett.} \textbf{\bibinfo{volume}{68}},
  \bibinfo{pages}{1943} (\bibinfo{year}{1992}).

\bibitem[{\citenamefont{Carra et~al.}(1993)\citenamefont{Carra, Thole,
  Altarelli, and Wang}}]{carra93}
\bibinfo{author}{\bibfnamefont{P.}~\bibnamefont{Carra}},
  \bibinfo{author}{\bibfnamefont{B.~T.} \bibnamefont{Thole}},
  \bibinfo{author}{\bibfnamefont{M.}~\bibnamefont{Altarelli}},
  \bibnamefont{and} \bibinfo{author}{\bibfnamefont{X.}~\bibnamefont{Wang}},
  \bibinfo{journal}{Phys. Rev. Lett.} \textbf{\bibinfo{volume}{70}},
  \bibinfo{pages}{694} (\bibinfo{year}{1993}).

\bibitem[{\citenamefont{Collins et~al.}(1995)\citenamefont{Collins, Laundy,
  Tang, and van~der Laan}}]{collins95}
\bibinfo{author}{\bibfnamefont{S.~P.} \bibnamefont{Collins}},
  \bibinfo{author}{\bibfnamefont{D.}~\bibnamefont{Laundy}},
  \bibinfo{author}{\bibfnamefont{C.~C.} \bibnamefont{Tang}}, \bibnamefont{and}
  \bibinfo{author}{\bibfnamefont{G.}~\bibnamefont{van~der Laan}},
  \bibinfo{journal}{J. Phys-Condens. Mat.} \textbf{\bibinfo{volume}{7}},
  \bibinfo{pages}{9325} (\bibinfo{year}{1995}).

\bibitem[{\citenamefont{Thole and van~der Laan}(1988)}]{thole88}
\bibinfo{author}{\bibfnamefont{B.~T.} \bibnamefont{Thole}} \bibnamefont{and}
  \bibinfo{author}{\bibfnamefont{G.}~\bibnamefont{van~der Laan}},
  \bibinfo{journal}{Phys. Rev. A} \textbf{\bibinfo{volume}{38}},
  \bibinfo{pages}{1943} (\bibinfo{year}{1988}).

\bibitem[{\citenamefont{Wilhelm et~al.}(2013)\citenamefont{Wilhelm, Eloirdi,
  Rusz, Springell, Colineau, Griveau, Oppeneer, Caciuffo, Rogalev, and
  Lander}}]{wilhelm13}
\bibinfo{author}{\bibfnamefont{F.}~\bibnamefont{Wilhelm}},
  \bibinfo{author}{\bibfnamefont{R.}~\bibnamefont{Eloirdi}},
  \bibinfo{author}{\bibfnamefont{J.}~\bibnamefont{Rusz}},
  \bibinfo{author}{\bibfnamefont{R.}~\bibnamefont{Springell}},
  \bibinfo{author}{\bibfnamefont{E.}~\bibnamefont{Colineau}},
  \bibinfo{author}{\bibfnamefont{J.-C.} \bibnamefont{Griveau}},
  \bibinfo{author}{\bibfnamefont{P.~M.} \bibnamefont{Oppeneer}},
  \bibinfo{author}{\bibfnamefont{R.}~\bibnamefont{Caciuffo}},
  \bibinfo{author}{\bibfnamefont{A.}~\bibnamefont{Rogalev}}, \bibnamefont{and}
  \bibinfo{author}{\bibfnamefont{G.~H.} \bibnamefont{Lander}},
  \bibinfo{journal}{Phys. Rev. B} \textbf{\bibinfo{volume}{88}},
  \bibinfo{pages}{024424} (\bibinfo{year}{2013}).

\bibitem[{\citenamefont{Shick et~al.}(1997)\citenamefont{Shick, Novikov, and
  Freeman}}]{shick97}
\bibinfo{author}{\bibfnamefont{A.~B.} \bibnamefont{Shick}},
  \bibinfo{author}{\bibfnamefont{D.~L.} \bibnamefont{Novikov}},
  \bibnamefont{and} \bibinfo{author}{\bibfnamefont{A.~J.}
  \bibnamefont{Freeman}}, \bibinfo{journal}{Phys. Rev. B}
  \textbf{\bibinfo{volume}{56}}, \bibinfo{pages}{14259(R)}
  (\bibinfo{year}{1997}).

\bibitem[{\citenamefont{Shick and Pickett}(2001)}]{shick01}
\bibinfo{author}{\bibfnamefont{A.~B.} \bibnamefont{Shick}} \bibnamefont{and}
  \bibinfo{author}{\bibfnamefont{W.~E.} \bibnamefont{Pickett}},
  \bibinfo{journal}{Phys. Rev. Lett.} \textbf{\bibinfo{volume}{86}},
  \bibinfo{pages}{300} (\bibinfo{year}{2001}).

\bibitem[{\citenamefont{Brooks et~al.}(1989)\citenamefont{Brooks, Eriksson, and
  Johansson}}]{brooks89}
\bibinfo{author}{\bibfnamefont{M.}~\bibnamefont{Brooks}},
  \bibinfo{author}{\bibfnamefont{O.}~\bibnamefont{Eriksson}}, \bibnamefont{and}
  \bibinfo{author}{\bibfnamefont{B.}~\bibnamefont{Johansson}},
  \bibinfo{journal}{J. Phys-Condens. Mat.} \textbf{\bibinfo{volume}{1}},
  \bibinfo{pages}{5861} (\bibinfo{year}{1989}).

\bibitem[{\citenamefont{Moore and van~der Laan}(2009)}]{moore09}
\bibinfo{author}{\bibfnamefont{K.~T.} \bibnamefont{Moore}} \bibnamefont{and}
  \bibinfo{author}{\bibfnamefont{G.}~\bibnamefont{van~der Laan}},
  \bibinfo{journal}{Rev. Mod. Phys.} \textbf{\bibinfo{volume}{81}},
  \bibinfo{pages}{235} (\bibinfo{year}{2009}).

\bibitem[{\citenamefont{Gignoux and Schmitt}(1997)}]{gignoux97}
\bibinfo{author}{\bibfnamefont{D.}~\bibnamefont{Gignoux}} \bibnamefont{and}
  \bibinfo{author}{\bibfnamefont{D.}~\bibnamefont{Schmitt}}, in
  \emph{\bibinfo{booktitle}{Handbook of Magnetic Materials}}, edited by
  \bibinfo{editor}{\bibfnamefont{K.~H.~W.} \bibnamefont{Buschow}}
  (\bibinfo{publisher}{North-Holland, Amsterdam}, \bibinfo{year}{1997}),
  vol.~\bibinfo{volume}{10}, p. \bibinfo{pages}{239}.

\bibitem[{\citenamefont{Moze et~al.}(1990)\citenamefont{Moze, Ibberson,
  Caciuffo, and Buschow}}]{caciuffo90}
\bibinfo{author}{\bibfnamefont{O.}~\bibnamefont{Moze}},
  \bibinfo{author}{\bibfnamefont{R.~M.} \bibnamefont{Ibberson}},
  \bibinfo{author}{\bibfnamefont{R.}~\bibnamefont{Caciuffo}}, \bibnamefont{and}
  \bibinfo{author}{\bibfnamefont{K.~H.~J.} \bibnamefont{Buschow}},
  \bibinfo{journal}{J. Less-Comm. Metals} \textbf{\bibinfo{volume}{166}},
  \bibinfo{pages}{329} (\bibinfo{year}{1990}).

\bibitem[{\citenamefont{Caciuffo et~al.}(1995)\citenamefont{Caciuffo, Amoretti,
  Buschow, Moze, Murani, and Paci}}]{caciuffo95}
\bibinfo{author}{\bibfnamefont{R.}~\bibnamefont{Caciuffo}},
  \bibinfo{author}{\bibfnamefont{G.}~\bibnamefont{Amoretti}},
  \bibinfo{author}{\bibfnamefont{K.~H.~J.} \bibnamefont{Buschow}},
  \bibinfo{author}{\bibfnamefont{O.}~\bibnamefont{Moze}},
  \bibinfo{author}{\bibfnamefont{A.~P.} \bibnamefont{Murani}},
  \bibnamefont{and} \bibinfo{author}{\bibfnamefont{B.}~\bibnamefont{Paci}},
  \bibinfo{journal}{J. Phys-Condens. Mat.} \textbf{\bibinfo{volume}{7}},
  \bibinfo{pages}{7981} (\bibinfo{year}{1995}).

\bibitem[{\citenamefont{Amoretti}(1984)}]{amoretti84}
\bibinfo{author}{\bibfnamefont{G.}~\bibnamefont{Amoretti}},
  \bibinfo{journal}{J. Physique} \textbf{\bibinfo{volume}{45}},
  \bibinfo{pages}{1067} (\bibinfo{year}{1984}).

\bibitem[{\citenamefont{van Steenwijk et~al.}(1977)\citenamefont{van Steenwijk,
  Lefever, Thiel, and Buschow}}]{vansteenwijk77}
\bibinfo{author}{\bibfnamefont{F.~J.} \bibnamefont{van Steenwijk}},
  \bibinfo{author}{\bibfnamefont{H.~T.} \bibnamefont{Lefever}},
  \bibinfo{author}{\bibfnamefont{R.~C.} \bibnamefont{Thiel}}, \bibnamefont{and}
  \bibinfo{author}{\bibfnamefont{K.~H.~J.} \bibnamefont{Buschow}},
  \bibinfo{journal}{Physica B+C} \textbf{\bibinfo{volume}{92}},
  \bibinfo{pages}{52 } (\bibinfo{year}{1977}).

\bibitem[{\citenamefont{Sanchez et~al.}(1988)\citenamefont{Sanchez, Burlet,
  Quezel, Bonnisseau, Rossat-Mignod, Spirlet, and Vogt}}]{sanchez88}
\bibinfo{author}{\bibfnamefont{J.-P.} \bibnamefont{Sanchez}},
  \bibinfo{author}{\bibfnamefont{P.}~\bibnamefont{Burlet}},
  \bibinfo{author}{\bibfnamefont{S.}~\bibnamefont{Quezel}},
  \bibinfo{author}{\bibfnamefont{D.}~\bibnamefont{Bonnisseau}},
  \bibinfo{author}{\bibfnamefont{J.}~\bibnamefont{Rossat-Mignod}},
  \bibinfo{author}{\bibfnamefont{J.-C.} \bibnamefont{Spirlet}},
  \bibnamefont{and} \bibinfo{author}{\bibfnamefont{O.}~\bibnamefont{Vogt}},
  \bibinfo{journal}{Solid State Commun.} \textbf{\bibinfo{volume}{67}},
  \bibinfo{pages}{999} (\bibinfo{year}{1988}).

\bibitem[{\citenamefont{Tanaka et~al.}(1992)\citenamefont{Tanaka, Laubacher,
  Steffen, Shera, Wohlfahrt, and Hoehn}}]{tanaka92}
\bibinfo{author}{\bibfnamefont{Y.}~\bibnamefont{Tanaka}},
  \bibinfo{author}{\bibfnamefont{D.~B.} \bibnamefont{Laubacher}},
  \bibinfo{author}{\bibfnamefont{R.~M.} \bibnamefont{Steffen}},
  \bibinfo{author}{\bibfnamefont{E.~B.} \bibnamefont{Shera}},
  \bibinfo{author}{\bibfnamefont{H.~D.} \bibnamefont{Wohlfahrt}},
  \bibnamefont{and} \bibinfo{author}{\bibfnamefont{M.~V.} \bibnamefont{Hoehn}},
  \bibinfo{journal}{Phys. Lett.} \textbf{\bibinfo{volume}{108 B}},
  \bibinfo{pages}{8} (\bibinfo{year}{1992}).

\bibitem[{\citenamefont{Barton and Cashion}(1979)}]{barton79}
\bibinfo{author}{\bibfnamefont{W.~A.} \bibnamefont{Barton}} \bibnamefont{and}
  \bibinfo{author}{\bibfnamefont{J.~D.} \bibnamefont{Cashion}},
  \bibinfo{journal}{J. Phys. C} \textbf{\bibinfo{volume}{12}},
  \bibinfo{pages}{2897} (\bibinfo{year}{1979}).

\bibitem[{\citenamefont{Thevenin and Cousson}(1987)}]{thevenin87}
\bibinfo{author}{\bibfnamefont{T.}~\bibnamefont{Thevenin}} \bibnamefont{and}
  \bibinfo{author}{\bibfnamefont{A.}~\bibnamefont{Cousson}},
  \bibinfo{journal}{J. Solid State Chem.} \textbf{\bibinfo{volume}{67}},
  \bibinfo{pages}{254} (\bibinfo{year}{1987}).

\bibitem[{\citenamefont{Kalvius et~al.}(1985)\citenamefont{Kalvius, Potzel,
  Moser, Litterst, Asch, Z\"{a}nkert, Potzel, Kratzer, Wunsch, Gal
  et~al.}}]{kalvius85}
\bibinfo{author}{\bibfnamefont{G.~M.} \bibnamefont{Kalvius}},
  \bibinfo{author}{\bibfnamefont{W.}~\bibnamefont{Potzel}},
  \bibinfo{author}{\bibfnamefont{J.}~\bibnamefont{Moser}},
  \bibinfo{author}{\bibfnamefont{F.~J.} \bibnamefont{Litterst}},
  \bibinfo{author}{\bibfnamefont{L.}~\bibnamefont{Asch}},
  \bibinfo{author}{\bibfnamefont{J.}~\bibnamefont{Z\"{a}nkert}},
  \bibinfo{author}{\bibfnamefont{U.}~\bibnamefont{Potzel}},
  \bibinfo{author}{\bibfnamefont{A.}~\bibnamefont{Kratzer}},
  \bibinfo{author}{\bibfnamefont{M.}~\bibnamefont{Wunsch}},
  \bibinfo{author}{\bibfnamefont{J.}~\bibnamefont{Gal}},
  \bibinfo{author}{\bibfnamefont{S.}~\bibnamefont{Fredo}},
  \bibinfo{author}{\bibfnamefont{D.}~\bibnamefont{Dayan}},
  \bibinfo{author}{\bibfnamefont{M. P.}~\bibnamefont{Dariel}},
  \bibinfo{author}{\bibfnamefont{M.}~\bibnamefont{Boge}},
  \bibinfo{author}{\bibfnamefont{J.}~\bibnamefont{Chappert}},
  \bibinfo{author}{\bibfnamefont{J. C.}~\bibnamefont{Spirlet}},
  \bibinfo{author}{\bibfnamefont{U.}~\bibnamefont{Benedict}}, \bibnamefont{and}
  \bibinfo{author}{\bibfnamefont{B. D.}~\bibnamefont{Dunlap}},
  \bibinfo{journal}{Physica B+C} \textbf{\bibinfo{volume}{130}},
  \bibinfo{pages}{393 } (\bibinfo{year}{1985}).

\bibitem[{\citenamefont{Gal et~al.}(1985)\citenamefont{Gal, Pinto, Fredo,
  Melamud, Shaked, Caciuffo, Litterst, Asch, Potzel, and Kalvius}}]{gal85}
\bibinfo{author}{\bibfnamefont{J.}~\bibnamefont{Gal}},
  \bibinfo{author}{\bibfnamefont{H.}~\bibnamefont{Pinto}},
  \bibinfo{author}{\bibfnamefont{S.}~\bibnamefont{Fredo}},
  \bibinfo{author}{\bibfnamefont{M.}~\bibnamefont{Melamud}},
  \bibinfo{author}{\bibfnamefont{H.}~\bibnamefont{Shaked}},
  \bibinfo{author}{\bibfnamefont{R.}~\bibnamefont{Caciuffo}},
  \bibinfo{author}{\bibfnamefont{F.~J.} \bibnamefont{Litterst}},
  \bibinfo{author}{\bibfnamefont{L.}~\bibnamefont{Asch}},
  \bibinfo{author}{\bibfnamefont{W.}~\bibnamefont{Potzel}}, \bibnamefont{and}
  \bibinfo{author}{\bibfnamefont{G.}~\bibnamefont{Kalvius}},
  \bibinfo{journal}{J. Magn. Magn. Mater.} \textbf{\bibinfo{volume}{50}},
  \bibinfo{pages}{L123 } (\bibinfo{year}{1985}).

\end{thebibliography}
\end{document}